\documentclass[aps,pre,amsmath,twocolumn,10pt,superscriptaddress,longbibliography]{revtex4-1}
\usepackage[dvipdfmx]{graphicx}
\usepackage{color}
\usepackage{bm}
\usepackage{braket}
\usepackage{amsmath,amssymb}
\usepackage{mathtools}

\usepackage{physics}
\usepackage[normalem]{ulem}

\begin{document}
\title{Projector-based efficient estimation of force constants}
\author{Atsuto \surname{Seko}}
\email{seko@cms.mtl.kyoto-u.ac.jp}
\affiliation{Department of Materials Science and Engineering, Kyoto University, Kyoto 606-8501, Japan}
\author{Atsushi \surname{Togo}}
\affiliation{Center for Basic Research on Materials National Institute for Materials Science, Tsukuba, Ibaraki 305-0047, Japan}

\date{\today}

\begin{abstract}
Estimating force constants for crystal structures is crucial for calculating various phonon-related properties. 
However, this task becomes particularly challenging when dealing with a large number of atoms or when third- and higher-order force constants are required.
In this study, we propose an efficient approach that involves constructing a complete orthonormal basis set for the force constants. 
We formulate this approach using projection matrices and their eigenvectors to meet the requirements for the force constants. 
This basis set enables precise inference of the force constants from displacement-force datasets.
Our efficient algorithms for basis-set construction and force constant estimation are implemented in the symfc code. 
Furthermore, several applications demonstrated in this study indicate that the current approach facilitates the efficient and accurate determination of force constants.
\end{abstract}

\maketitle

\section{Introduction}

Force constants in crystal structures are essential for calculating various phonon-related properties \cite{IntroductionToLatticeDynamics, Thermodynamics-of-crystals, Ziman-electrons-phonons}.
To estimate these force constants, supercell approaches combined with the finite displacement method are commonly employed \cite{Parlinski-phonon-1997, Hellman-TDEP-2013, Tadano-ALM-2018, hiPhive, phonopy-phono3py-JPSJ}.
Additionally, in self-consistent phonon approaches, effective force constants are often derived from supercell structures that replicate the atomic distribution at specific temperatures \cite{Errea-SSCHA-2013, van-Roekeghem-2020}.
These methods involve introducing random or systematic atomic displacements into supercells, calculating the forces on the atoms, and then estimating the supercell force constants from the resulting displacement-force dataset.

Determining the force-constant elements requires numerous supercells with different displacement configurations, because the number of elements to be determined can be substantial. 
At the same time, the number of force-constant elements can be reduced by using a basis set that represents the force constants, applying permutation symmetry rules, which stem from the definition of force constants as derivatives of the potential energy, and sum rules, which derive from the invariance under infinitesimal translations \cite{Thermodynamics-of-crystals}. 
Additionally, the symmetric properties of the structure can further reduce the number of required force-constant elements \cite{Tadano-ALM-2018, hiPhive, Physical-Properties-of-Crystals}. 
A significantly reduced but complete orthonormal basis set enables efficient and accurate calculation of force constants that satisfy necessary requirements \cite{el-batanouny_wooten_2008, Dynamics-of-perfect-crystals}.
However, constructing such a basis set and estimating the force constants using the basis set can be computationally intensive, particularly for systems with many atoms or when higher-order force constants are involved.

Various approaches, including those utilizing compressed sensing, have been developed to estimate force constants with high precision \cite{Zhou-PRL-compressive-sensing-FC-2014, Tadano-2015}. 
Despite the availability of other approaches \cite{Tadano-ALM-2018, hiPhive, Hellman-TDEP-2013}, force constant estimation remains a computationally challenging task. 
In this study, we propose a projector-based approach that efficiently constructs a complete orthonormal basis set for force constants. 
We introduce several compression techniques that allow for a reduction in the size of the basis set.
These techniques involve compressing a projection matrix using eigenvectors of another projection matrix onto a subspace defined by partial requirements for the force constants.
The current procedures for efficient basis-set construction and force constant estimation from displacement-force datasets are implemented in the symfc code \cite{symfc-project}. We demonstrate various applications of this approach, showing that it enables the efficient and precise determination of force constants.

Section \ref{symfc:sec-projection-matrix-formulation} introduces a straightforward projector-based formulation for constructing basis sets for force constants. 
Section \ref{symfc:sec-efficient-basis-set-construction} presents various techniques for efficiently constructing basis sets using projection matrices, enabling the construction of basis sets for higher-order force constants.
Section \ref{symfc:sec-procedure-basis-set} outlines the current procedure for constructing basis sets that exactly satisfy the required constraints for force constants.
Section \ref{symfc:sec-force-constant-estimation} describes the procedure for estimating force constants from a displacement-force dataset using the constructed basis set.
Section \ref{symfc:sec-results} demonstrates several applications of the proposed projector-based procedures for basis set construction and force constant estimation in diamond silicon and ionic systems. 
Finally, Sec. \ref{symfc:sec-conclusion} provides the concluding remarks.

\section{Projection matrices for force constants}
\label{symfc:sec-projection-matrix-formulation}
\subsection{Crystal potential and force constants}

Force constants of a crystal are derived from the Taylor expansion of the potential energy $\mathcal{V}$ with respect to atomic displacements.
In this study, we consider supercell force constants, which are estimated from a set of finite displacements and forces acting on atoms in supercells.
For a supercell with $N$ atoms, the potential energy is represented by 
\begin{eqnarray}
\mathcal{V} &=& \Theta_0 + \sum_{i\alpha} \Theta_{i\alpha} u_{i\alpha} 
+ \frac{1}{2} \sum_{i\alpha,j\beta} \Theta_{i\alpha,j\beta} u_{i\alpha} u_{j\beta} \nonumber \\
&+& \frac{1}{3!} \sum_{i\alpha,j\beta,k\gamma} \Theta_{i\alpha,j\beta,k\gamma} 
u_{i\alpha} u_{j\beta} u_{k\gamma} + \cdots,
\end{eqnarray}
%\begin{eqnarray}
%\mathcal{V} = \Theta_0 &+& \sum_{i,\alpha} \Theta_{i\alpha} u_{i\alpha} \nonumber\\
%&+& \frac{1}{2} \sum_{i\alpha,j\beta} \Theta_{i\alpha,j\beta} u_{i\alpha} u_{j\beta} \\
%&+& \frac{1}{3!} \sum_{i\alpha,j\beta,k\gamma} \Theta_{i\alpha,j\beta,k\gamma} 
%u_{i\alpha} u_{j\beta} u_{k\gamma} \nonumber \\
%&+& \cdots \nonumber,
%\end{eqnarray}
where $i$, $j$, and $k$ are the atomic indices within the supercell, and $\alpha$, $\beta$, and $\gamma$ are the Cartesian components. 
The second-order and third-order supercell force constants are represented by $\Theta_{i\alpha,j\beta}$ and $\Theta_{i\alpha,j\beta,k\gamma}$, respectively. 
In addition, we assume that the first-order force constants $\Theta_{i\alpha}$ is equal to zero because atomic displacements are measured from their equilibrium positions.

\subsection{Projection matrix for second-order force constants}
In the current formulation, the second-order force constants are represented as a column vector with $9N^2$ elements, denoted as 
\begin{equation}
\bm{\theta}^{\rm (FC2)} = [\Theta_{1x,1x}, \Theta_{1x,1y}, \Theta_{1x,1z}, \Theta_{1x,2x}, \cdots]^\top, 
\end{equation}
where the components \(\Theta_{i\alpha,j\beta}\) are listed in ascending order of the indices \((i, \alpha, j, \beta)\).
We start by considering the force constants that are invariant with the space group operations $\mathcal{G}$ for the target structure. 
Since we are discussing supercell force constants, the lattice translation subgroup in $\mathcal{G}$ is limited to those of the lattice points inside the supercell.
An operation $\hat g \in \mathcal{G}$ acts on the force constant vector as
\begin{equation}
\hat g \bm{\theta}^{\rm (FC2)} = \bm{\theta}^{\rm (FC2)} \left[ \bm{\Gamma} (\hat g) \otimes \bm{\Gamma} (\hat g) \right], 
\end{equation}
where $\bm{\Gamma} (\hat g)$ denotes the matrix representation of operation $\hat g$ for the first-order force constants $\{\Theta_{i\alpha}\}$.
We use the group-theoretical projection operator method \cite{el-batanouny_wooten_2008} to reduce the Kronecker product of representations to the identity irreducible representation.
The projection matrix for the identity irreducible representation is given as
\begin{equation}
\bm{P}_{\rm spg} = \frac{1}{|\mathcal{G}|} \sum_{\hat g \in \mathcal{G}} \bm{\Gamma} (\hat g) \otimes \bm{\Gamma} (\hat g),
\end{equation}
where $|\mathcal{G}|$ denotes the order of group $\mathcal{G}$.

Then, we introduce projection matrices to enforce the constraints associated with the sum rules and permutation symmetry rules for the force constants. 
The sum rule for $(i\alpha,\beta)$ and the permutation symmetry rule for $(i\alpha,j\beta)$ are expressed as 
\begin{equation}
\label{symfc:eq-sumrule}
\sum_j \Theta_{i\alpha,j\beta} = 0
\end{equation}
and
\begin{equation}
\label{symfc:eq-permutation}
\Theta_{i\alpha,j\beta} - \Theta_{j\beta,i\alpha} = 0,
\end{equation}
respectively. 
These rules impose linear constraints on the force constants, which are mutually independent. 
In matrix form, these constraints are represented as $\bm{C}^\top\bm{\theta}^{\rm (FC2)} = 0$. 
Therefore, a force-constant basis set that satisfies these linear constraints spans the kernel of $\bm{C}^\top$, which can be expressed as
\begin{equation} 
{\rm Ker} (\bm{C}^\top) = \{\bm{\theta}^{\rm (FC2)} \in \mathbb{R}^{9N^2} | \bm{C}^\top\bm{\theta}^{\rm (FC2)} = 0 \}.
\end{equation}
Since ${\rm Ker} (\bm{C}^\top)$ is the complementary subspace spanned by $\bm{C}$, the orthogonal projection matrix onto ${\rm Ker} (\bm{C}^\top)$ is given by \cite{Strang_2023,Yanai:1414711}
\begin{equation}
\label{symfc:Eqn-projector-constraints}
\bm{P}_{\rm perm \cap sum} = \bm{I} - \bm{C} (\bm{C^\top C}) ^{-1} \bm{C}^\top,
\end{equation}
where $\bm{I}$ denotes the unit matrix of size $9N^2$.
A derivation of the projection matrix can be found in Appendix \ref{symfc:sec:derivation-of-projector}.

Each column of $\bm{C}$ corresponds to either the sum rule or the permutation symmetry rule for a specific combination of indices.
The sum rule for $(i,\alpha,\beta)$ can be expressed using the column unit vector $\bm{v}_{i\alpha\beta}$ such that $\bm{v}_{i\alpha\beta}^\top \bm{\theta}^{\rm (FC2)} = 0$.
The vector $\bm{v}_{i\alpha\beta}$ contains $N$ non-zero elements, and the set of vectors $\{\bm{v}_{i\alpha\beta}\}$ is orthonormal.
Similarly, the permutation symmetry rule for the combination $(i\alpha, j\beta)$ is described by the column unit vector $\bm{w}_{i\alpha,j\beta}$ such that $\bm{w}_{i\alpha,j\beta}^\top \bm{\theta}^{\rm (FC2)} = 0$.
The vector $\bm{w}_{i\alpha,j\beta}$ has only two non-zero elements: $1/\sqrt{2}$ and $-1/\sqrt{2}$.
Additionally, the set of vectors $\{\bm{w}_{i\alpha,j\beta}\}$ is orthonormal if the index combinations are distinct and do not contain duplicates.

Using a complete set of $\bm{v}_{i\alpha\beta}$ and $\bm{w}_{i\alpha,j\beta}$, the matrix $\bm{C}$ is described as
\begin{equation}
\bm{C} = [\bm{C}_v, \bm{C}_w],
\end{equation}
where
\begin{equation}
\bm{C}_v = [\cdots, \bm{v}_{i\alpha\beta}, \cdots], \:\: \bm{C}_w = [\cdots, \bm{w}_{i\alpha,j\beta}, \cdots].
\end{equation}
The matrices $\bm{C}_v$ and $\bm{C}_w$ have dimensions $(9N^2, 9N)$ and approximately $(9N^2, 9N^2/2)$, respectively.
Since the vectors $\bm{v}_{i\alpha\beta}$ and $\bm{w}_{i\alpha,j\beta}$ are linearly independent but not necessarily orthogonal, the projection matrix onto ${\rm Ker} (\bm{C}^\top)$ is given by Eq. (\ref{symfc:Eqn-projector-constraints}), which involves the inverse of $\bm{C^\top C}$.

Finally, we present the full projection matrix onto the intersection of vector subspaces determined by projection matrices $\bm{P}_{\rm spg}$ and $\bm{P}_{\rm perm \cap sum}$.
Since projection matrices $\bm{P}_{\rm spg}$ and $\bm{P}_{\rm perm \cap sum}$ are commutative, as shown in Appendix \ref{symfc:sec:commutativity-proof}, the full projection matrix can be written as $\bm{P} = \bm{P}_{\rm spg} \bm{P}_{\rm perm \cap sum}$ \cite{Yanai:1414711,PIZIAK199967}.
By solving the eigenvalue problem for the full projection matrix:
\begin{equation}
\bm{P} \bm{b} = \bm{b},
\end{equation}
we can obtain an orthonormal basis set for representing the force constants, $\bm{B}$.
This basis set is invariant under space group operations and satisfies both the sum rules and the permutation symmetry rules. 
Since all non-zero eigenvalues of the projection matrix are equal to one, the number of independent basis vectors corresponds to the trace of the projection matrix.

\subsection{Projection matrix for third-order force constants}

The third-order force constants are represented in a vector form with $27N^3$ elements as 
\begin{equation}
\bm{\theta}^{\rm (FC3)} = [\Theta_{1x,1x,1x}, \Theta_{1x,1x,1y}, \Theta_{1x,1x,1z}, \Theta_{1x,1x,2x}, \cdots]^\top, 
\end{equation}
which are listed in ascending order of $(i,\alpha,j,\beta,k,\gamma)$. 
An operation $\hat g \in \mathcal{G}$ acts on the third-order force constant vector as
\begin{equation}
\hat g \bm{\theta}^{\rm (FC3)} = \bm{\theta}^{\rm (FC3)} \left[ \bm{\Gamma} (\hat g) \otimes \bm{\Gamma} (\hat g) \otimes \bm{\Gamma} (\hat g) \right].
\end{equation}
Therefore, the projection matrix for the identity irreducible representation is given as
\begin{equation}
\bm{P}_{\rm spg} = \frac{1}{|\mathcal{G}|} \sum_{\hat g \in \mathcal{G}} \bm{\Gamma} (\hat g) \otimes \bm{\Gamma} (\hat g) \otimes \bm{\Gamma} (\hat g).
\end{equation}
The sum rules and permutation symmetry rules can also be formulated in matrix form as $\bm{C}^\top\bm{\theta}^{(\rm FC3)} = 0$, analogous to the matrix representation used for second-order force constants.
The kernel of $\bm{C}^\top$ is written as
\begin{equation} 
{\rm Ker} (\bm{C}^\top) = \{\bm{\theta}^{(\rm FC3)} \in \mathbb{R}^{27N^3} | \bm{C}^\top\bm{\theta}^{(\rm FC3)} = 0 \},
\end{equation}
and the orthogonal projection matrix onto ${\rm Ker} (\bm{C}^\top)$ is given by \cite{Strang_2023,Yanai:1414711}
\begin{equation}
\label{symfc:Eqn-projector-constraints-fc3}
\bm{P}_{\rm perm \cap sum} = \bm{I} - \bm{C} (\bm{C^\top C}) ^{-1} \bm{C}^\top,
\end{equation}
where $\bm{I}$ denotes the unit matrix of size $27N^3$.

The sum rule for $(i,\alpha,j,\beta,\gamma)$ is given by
\begin{equation}
\label{symfc:eq-sumrule-fc3}
\sum_k \Theta_{i\alpha,j\beta,k\gamma} = 0,
\end{equation}
and can be expressed as $\bm{v}_{i{\alpha}j\beta\gamma}^\top \bm{\theta}^{(\rm FC3)} = 0$.
The vector $\bm{v}_{i{\alpha}j\beta\gamma}$ contains only $N$ non-zero elements out of a total of $27N^3$ elements, and the set of vectors $\{\bm{v}_{i{\alpha}j\beta\gamma}\}$ is mutually orthonormal.

Representing the permutation symmetry rules for third-order force constants is more complex than for second-order force constants.
For the permutation symmetry of $(i\alpha,j\beta,k\gamma)$, the rules are expressed as
\begin{eqnarray}
\label{symfc:eq-permutation-fc3}
\Theta_{i\alpha,j\beta,k\gamma}-\Theta_{i\alpha,k\gamma,j\beta} &=& 0 \nonumber \\
\Theta_{i\alpha,j\beta,k\gamma}-\Theta_{j\beta,i\alpha,k\gamma} &=& 0 \nonumber \\
\Theta_{i\alpha,j\beta,k\gamma}-\Theta_{j\beta,k\gamma,i\alpha} &=& 0 \\
\Theta_{i\alpha,j\beta,k\gamma}-\Theta_{k\gamma,i\alpha,j\beta} &=& 0 \nonumber \\
\Theta_{i\alpha,j\beta,k\gamma}-\Theta_{k\gamma,j\beta,i\alpha} &=& 0, \nonumber  
\end{eqnarray}
when all indices $i\alpha$, $j\beta$, and $k\gamma$ are distinct.
These constraints can be represented using a matrix composed of column unit vectors $\bm{W}_{i\alpha,j\beta,k\gamma}$ such that $\bm{W}_{i\alpha,j\beta,k\gamma}^\top \bm{\theta}^{(\rm FC3)} = 0$. 
Each column vector of $\bm{W}_{i\alpha,j\beta,k\gamma}$ contains two non-zero elements, $1/\sqrt{2}$ and $-1/\sqrt{2}$. 
Although these vectors are not orthogonal, they are linearly independent.
For the case where the indices are of the form $(i\alpha,i\alpha,k\gamma)$, the permutation symmetry rules are given by
\begin{eqnarray}
\label{symfc:eq-permutation-fc3-2}
\Theta_{i\alpha,i\alpha,k\gamma}-\Theta_{i\alpha,k\gamma,i\alpha} &=& 0 \nonumber \\
\Theta_{i\alpha,i\alpha,k\gamma}-\Theta_{k\gamma,i\alpha,i\alpha} &=& 0.
\end{eqnarray}
Similarly, the matrix of column unit vectors $\bm{W}_{i\alpha,i\alpha,k\gamma}$ is defined such that $\bm{W}_{i\alpha,i\alpha,k\gamma}^\top \bm{\theta}^{(\rm FC3)} = 0$. 
Each column vector in $\bm{W}_{i\alpha,i\alpha,k\gamma}$ has two non-zero elements, $1/\sqrt{2}$ and $-1/\sqrt{2}$.

Using a linearly independent complete set of $\bm{v}_{i{\alpha}j\beta\gamma}$ and $\bm{W}_{i\alpha,j\beta,k\gamma}$, the matrix $\bm{C}$ is described as
\begin{equation}
\bm{C} = [\bm{C}_v, \bm{C}_w],
\end{equation}
where
\begin{equation}
\bm{C}_v = [\cdots, \bm{v}_{i{\alpha}j\beta\gamma}, \cdots], 
\:\: \bm{C}_w = [\cdots, \bm{W}_{i\alpha,j\beta,k\gamma}, \cdots].
\end{equation}
The matrices $\bm{C}_v$ and $\bm{C}_w$ have dimensions $(27N^3, 27N^2)$ and approximately $(27N^3, (5/6)27N^3)$, respectively.
Finally, the full projection matrix is given by $\bm{P} = \bm{P}_{\rm spg} \bm{P}_{\rm perm \cap sum}$.
By solving the eigenvalue problem for the full projection matrix, $\bm{P} \bm{b} = \bm{b}$,
an orthonormal basis set for representing third-order force constants is obtained.
Note that the derivation of basis sets for representing higher-order force constants can be similarly formalized.
The symfc code implements the construction of basis sets up to fourth-order force constants.

\section{Efficient basis set construction}
\label{symfc:sec-efficient-basis-set-construction}

The previous section outlines a straightforward formulation of the projector-based approach for constructing a complete basis set for force constants. 
However, a direct application of this method requires computing the inverse of $\bm{C}^\top \bm{C}$ and solving large eigenvalue problems. 
Specifically, for second-order force constants, the eigenvalue problem is of size $(9N^2, 9N^2)$, and for third-order force constants, it is of size $(27N^3, 27N^3)$.
Such computations can be prohibitively expensive, particularly for the third-order force constants.
In Sec. \ref{symfc:sec-decomposition-projection-matrix}--\ref{symfc:sec-eigenvalue-problem}, we introduce and formulate various techniques designed to efficiently construct a basis set for force constants, thereby addressing associated computational challenges.
The overall procedure for constructing a basis set is outlined in Sec. \ref{symfc:sec-procedure-basis-set}.

\subsection{Decomposition of projection matrix}
\label{symfc:sec-decomposition-projection-matrix}
Instead of deriving the full projection matrix by computing the inverse of $\bm{C}^\top \bm{C}$, we construct a complete basis set using an approximation of the full projection matrix. 
This approximated projection matrix is composed of individual projection matrices that apply the sum rules and permutation symmetry rules separately.
For the second-order force constants, orthogonal projection matrices $\bm{P}_{\bm{v}}$ and $\bm{P}_{\bm{w}}$ onto ${\rm Ker} (\bm{C}_v^\top)$ and ${\rm Ker} (\bm{C}_w^\top)$ are expressed using orthonormal vector sets $\{\bm{v}_{i\alpha\beta}\}$ and $\{\bm{w}_{i\alpha,j\beta}\}$ as \cite{Strang_2023}
\begin{eqnarray}
\label{symfc:eqn:sum_rules}
\bm{P}_v &=& \bm{I} - \sum_{i=1}^N \sum_{\alpha=1}^3 \sum_{\beta=1}^3 \bm{v}_{i\alpha\beta} \bm{v}_{i\alpha\beta}^\top \nonumber \\
&=& \bm{I} - \bm{C}_v \bm{C}_v^\top
\end{eqnarray}
and
\begin{eqnarray}
\label{symfc:eqn:permutation_rules}
\bm{P}_w &=& \bm{I} - \sum_{(i\alpha,j\beta)} \bm{w}_{i\alpha,j\beta} \bm{w}_{i\alpha,j\beta}^\top \nonumber \\
&=& \bm{I} - \bm{C}_w \bm{C}_w^\top,
\end{eqnarray}
respectively.
For the third-order force constants, orthogonal projection matrices $\bm{P}_{\bm{v}}$ and $\bm{P}_{\bm{w}}$ onto ${\rm Ker} (\bm{C}_v^\top)$ and ${\rm Ker} (\bm{C}_w^\top)$ are given as
\begin{eqnarray}
\label{symfc:eqn:sum_rules-fc3}
\bm{P}_v &=& \bm{I} - \sum_{i=1}^N \sum_{\alpha=1}^3 \sum_{j=1}^N \sum_{\beta=1}^3 \sum_{\gamma=1}^3 \bm{v}_{i\alpha j\beta\gamma} \bm{v}_{i\alpha j\beta\gamma}^\top \nonumber \\
&=& \bm{I} - \bm{C}_v \bm{C}_v^\top
\end{eqnarray}
and
\begin{eqnarray}
\label{symfc:eqn:permutation_rules-fc3}
\bm{P}_w &=& \bm{I} - \bm{C}_w (\bm{C}_w^\top \bm{C}_w)^{-1} \bm{C}_w^\top,
\end{eqnarray}
respectively.

Thus, projection matrices that apply the sum rules and permutation symmetry rules can be defined individually.
However, the projection matrix $\bm{P}_{\rm perm \cap sum}$ is not simply the product of $\bm{P}_w$ and $\bm{P}_v$ due to the non-commutative nature of these matrices.
The projection matrix $\bm{P}_{\rm perm \cap sum}$ along with its eigenvectors will be derived as demonstrated in Sec. \ref{symfc:sec-non-commutative-matrices}.

\subsection{Non-commutative projection matrices}
\label{symfc:sec-non-commutative-matrices}

Given non-commutative projection matrices $\bm{P}_1$ and $\bm{P}_2$, the projection matrix $\bm{P}_{1 \cap 2}$ can be expressed using the von Neumann--Halperin theorem as \cite{Neumann1949OnRO,halperin1962product,PIZIAK199967,galantai2013projectors}
\begin{equation}
\label{symfc:eq-Neumann-Halperin}
\bm{P}_{1 \cap 2} = \lim_{p \to \infty} \bm{P}_{1} (\bm{P}_{2} \bm{P}_{1})^p,
\end{equation}
which represents the projection matrix onto the intersection of the subspaces associated with $\bm{P}_1$ and $\bm{P}_2$.
The geometric interpretation of Eq. (\ref{symfc:eq-Neumann-Halperin}) is that a vector \( \bm{x} \) is first projected onto the subspace associated with \( \bm{P}_{1} \), and the resulting vector is subsequently projected onto the subspace associated with \( \bm{P}_{2} \). 
This process of alternating projections onto the subspaces associated with \( \bm{P}_{1} \) and \( \bm{P}_{2} \) is repeated. 
In the end, the projected vector converges to \( \bm{P}_{1 \cap 2} \bm{x} \).

For practical purposes, $\bm{P}_{1 \cap 2}$ can be approximated using a finite value of $p$.
In this study, setting $p=1$ and approximating $\bm{P}_{1 \cap 2}$ as $\tilde{\bm{P}}_{1 \cap 2} \simeq \bm{P}_1 (\bm{P}_2 \bm{P}_1)^{1}$ suffice to derive a complete basis set for $\bm{P}_{1 \cap 2}$.
By solving the eigenvalue problem for the approximated projection matrix defined as
\begin{equation}
\label{symfc:eq-approx-proj-12}
\tilde{\bm{P}}_{1 \cap 2} = \bm{P}_1 \bm{P}_2 \bm{P}_1
\end{equation}
and removing the eigenvectors with eigenvalues less than one, a complete basis set for $\bm{P}_{1 \cap 2}$ can be obtained.
This can be derived from the fact that $\bm{P}_{1 \cap 2} = \lim_{p \to \infty} \tilde{\bm{P}}_{1 \cap 2}^p$.
Note that the matrix described by Eq. (\ref{symfc:eq-approx-proj-12}) is an approximation and does not fully satisfy the conditions of a projection matrix. 
Nonetheless, the basis set formed by its eigenvectors corresponding to eigenvalues of one constitutes a complete set of eigenvectors for the projection matrix $\bm{P}_{1 \cap 2}$.

For the non-commutative matrices $\bm{P}_v$ and $\bm{P}_w$, the projection matrix $\bm{P}_{\rm perm \cap sum}$ and the full projection matrix $\bm{P} = \bm{P}_{\rm spg} \bm{P}_{\rm perm \cap sum}$ can be approximated as
\begin{equation}
\label{symfc:eq-approx-proj-perm-sum}
\tilde{\bm{P}}_{\rm perm \cap sum} = \bm{P}_{w} \bm{P}_{v} \bm{P}_{w}
\end{equation}
and 
\begin{equation}
\label{symfc:eq-approx-proj-full}
\tilde{\bm{P}} = \bm{P}_{\rm spg} \bm{P}_{w} \bm{P}_{v} \bm{P}_{w},
\end{equation}
respectively.
Solving the eigenvalue problem for the approximated full projection matrix and removing the eigenvectors with eigenvalues less than one yields a complete basis set for the full projection matrix.

\subsection{Compression of projection matrices}
The compression of projection matrices is crucial for constructing a complete force-constant basis set of the full projection matrix without relying on approximations.
Although the complete basis set can be derived by solving the eigenvalue problem associated with the product of partial projection matrices given by Eq. (\ref{symfc:eq-approx-proj-full}), the computational demands of full-sized large projection matrices for third-order force constants, scaling as \((27N^3, 27N^3)\), and higher-order force constants present significant challenges for several reasons:
(1) The construction of full-sized partial projection matrices is computationally intensive and necessitates substantial memory allocation.
(2) The computation of products of full-sized partial projection matrices is also computationally demanding.
(3) Solving eigenvalue problems for full-sized projection matrices is prohibitively expensive in terms of computational resources.

Therefore, we propose a technique to compress projection matrices, enabling the execution of these operations with the smallest possible compressed projection matrices. 
This compression procedure is designed to retain all force-constant components essential for constructing the complete set of the full projection matrix.
It relies on the eigenvalue decompositions of partial projection matrices, where the eigenvectors can significantly enhance the compression of large matrices, provided that the eigenvectors of the partial projection matrix are well-compressed, sparse, and efficiently computed. 
In the following, we will demonstrate how to compress a projection matrix using the eigenvectors of a partial projection matrix.

Firstly, we begin with the eigenvalue decomposition of a partial projection matrix.
In this decomposition, a partial projection matrix \(\bm{P}_j\) is expressed in terms of its complete and orthonormal set of eigenvectors \(\bm{B}_j\) as 
\begin{equation}
\bm{P}_j = \bm{B}_j \bm{B}_j^\top,
\end{equation}
where \(\bm{B}_j\) has dimensions \((9N^2, n_j)\) for second-order force constants and \((27N^3, n_j)\) for third-order force constants. 
The variable \(n_j\) denotes the number of columns, corresponding to the size of the complete set of eigenvectors for \(\bm{P}_j\). 
Since the eigenvectors \(\bm{B}_j\) are well-compressed for the partial projection matrices considered in this study, meaning that the size of the complete set of eigenvectors is significantly smaller than the number of rows, they can be utilized for compressing other projection matrices.

The matrix \(\bm{P}_i\) can be effectively compressed using the eigenvectors of \(\bm{P}_j\), denoted as \(\bm{B}_j\), when the product \(\bm{P}_j \bm{P}_i \bm{P}_j\) can be represented as a component of the full projection matrix. 
This compression is applicable when the eigenvectors are either known or readily computable.
The compression process can be described as follows:
\begin{eqnarray}
\bm{P}_j\bm{P}_i\bm{P}_j 
& = &\bm{B}_j\bm{B}_j^\top \bm{P}_i \bm{B}_j\bm{B}_j^\top 
\:\:\: (\bm{P}_j = \bm{B}_j\bm{B}_j^\top) \nonumber\\
& = &\bm{B}_j \bm{P}_{i\downarrow j} \bm{B}_j^\top,
\end{eqnarray}
where $\bm{P}_{i\downarrow j}$ represents the compressed version of $\bm{P}_i$ in the basis defined by $\bm{B}_j$. 
It is defined as
\begin{equation}
\label{symfc:matrix-comp-definition}
\bm{P}_{i\downarrow j} = \bm{B}_j^\top \bm{P}_i \bm{B}_j.
\end{equation}
The size of $\bm{P}_{i\downarrow j}$ is $(n_j, n_j)$, which is significantly smaller than the size of $\bm{P}_i$.
It is important to note that a complete set of eigenvectors \(\bm{B}_j\) is employed to derive \(\bm{P}_{i \downarrow j}\), ensuring that this compression is exact and does not omit any basis vectors relevant to $\bm{P}_j\bm{P}_i\bm{P}_j$.

The eigenvectors of \(\bm{P}_j \bm{P}_i \bm{P}_j\) can be efficiently computed by solving the eigenvalue problem for the smaller matrix \(\bm{P}_{i \downarrow j}\) as follows.
(1) If the eigenvectors of $\bm{P}_i = \bm{B}_i\bm{B}_i^\top$ are known without explicitly constructing $\bm{P}_i$ and without solving its eigenvalue problem, then $\bm{P}_{i\downarrow j}$ can be written as 
\begin{equation}
\label{symfc:matrix-comp-definition2}
\bm{P}_{i\downarrow j} = \bm{B}_j^\top \bm{B}_i \bm{B}_i^\top \bm{B}_j.
\end{equation}
Since $\bm{B}_j^\top \bm{B}_i$ is generally smaller in size, we first compute $\bm{A} = \bm{B}_j^\top \bm{B}_i$, and then find $\bm{P}_{i\downarrow j}$ by calculating $\bm{P}_{i\downarrow j} = \bm{A} \bm{A}^\top$.
If $\bm{P}_i$ must be constructed, $\bm{P}_{i\downarrow j}$ can be computed using Eq. (\ref{symfc:matrix-comp-definition}).
(2) The eigenvalue problem for the compressed projection matrix $\bm{P}_{i\downarrow j}$ is solved as $\bm{P}_{i\downarrow j} = \bm{B}_{i\downarrow j}\bm{B}_{i\downarrow j}^\top$, where $\bm{B}_{i\downarrow j}$ are the compressed eigenvectors by $\bm{B}_j$.
(3) The eigenvectors of $\bm{P}_j\bm{P}_i\bm{P}_j$ with eigenvalues of one, $\bm{B}_{i \cap j}$, are determined as
\begin{equation}
\bm{B}_{i \cap j} = \bm{B}_j \bm{B}_{i\downarrow j},
\end{equation}
because $\bm{P}_j\bm{P}_i\bm{P}_j$ can be represented by 
\begin{eqnarray}
\bm{P}_j\bm{P}_i\bm{P}_j &=& \bm{B}_{i \cap j} \bm{B}_{i \cap j}^\top \nonumber \\
 &=& \bm{B}_j \bm{B}_{i\downarrow j} \bm{B}_{i\downarrow j}^\top \bm{B}_j^\top.
\end{eqnarray}
Thus, a complete set of eigenvectors of $\bm{P}_j\bm{P}_i\bm{P}_j$ can be calculated efficiently without the need to directly compute the product $\bm{P}_j\bm{P}_i\bm{P}_j$.

If the projection matrices $\bm{P}_i$ and $\bm{P}_j$ commute, i.e., $[\bm{P}_i,\bm{P}_j]=0$, this compression technique can be applied to the product $\bm{P}_i\bm{P}_j$.
Specifically, the product $\bm{P}_i\bm{P}_j$ can be expressed as
\begin{equation}
\bm{P}_i\bm{P}_j
 = \bm{P}_i\bm{P}_j^2
 = \bm{P}_j\bm{P}_i\bm{P}_j ,
\end{equation}
where we use the fact that projection matrices are idempotent, $\bm{P}_j^2 = \bm{P}_j$.

\subsection{Application of sum rules}
We utilize the equation $\bm{P}_v = \bm{I} - \bm{C}_v \bm{C}_v^\top$, which is presented in Eqs. (\ref{symfc:eqn:sum_rules}) and (\ref{symfc:eqn:sum_rules-fc3}), to incorporate the sum rules. 
The use of this formulation is computationally efficient because both the identity matrix $\bm{I}$ and the matrix $\bm{C}_v$ are sparse. 
On the other hand, the eigenvectors of $\bm{P}_v$, denoted by $\bm{B}_v$, typically contain a larger number of non-zero elements compared to $\bm{C}_v$.

\subsection{Application of permutation symmetry rules}
The eigenvectors of the projection matrix $\bm{P}_w$, denoted by $\bm{B}_w$, can be efficiently determined without the need to explicitly construct $\bm{P}_w$.
These eigenvectors span the vector space that satisfies the permutation symmetry rules.
For second-order force constants, the permutation symmetry rule for $(i\alpha,j\beta)$ is specified by Eq. (\ref{symfc:eqn:permutation_rules}) and is represented by a single vector $\bm{w}_{i\alpha,j\beta}$ such that $\bm{w}_{i\alpha,j\beta}^\top \bm{\theta}=0$.
Since the set of vectors $\{\bm{w}_{i\alpha,j\beta}\}$ is orthonormal, the unit vector $\bm{w}_{i\alpha,j\beta}^\perp$, which is orthogonal to $\bm{w}_{i\alpha,j\beta}$, can serve as an eigenvector of $\bm{P}_w$. 
This vector $\bm{w}_{i\alpha,j\beta}^\perp$ has only two non-zero elements of $1/\sqrt{2}$ corresponding to the columns $(i,\alpha,j,\beta)$ and $(j,\beta,i,\alpha)$.
In cases where $i\alpha=j\beta$, no permutation symmetry rules are applicable.
Therefore, $\bm{w}_{i\alpha,i\alpha}^\perp$ is defined as the vector with a single non-zero element of one for the column $(i,\alpha,i,\alpha)$.
Thus, the eigenvectors $\bm{B}_w$ are given by
\begin{equation}
\bm{B}_w = [\cdots, \bm{w}_{i\alpha,j\beta}^\perp, \cdots],
\end{equation}
where the vectors $\{\bm{w}_{i\alpha,j\beta}^\perp\}$ are mutually orthonormal.
The matrix $\bm{B}_w$ has dimensions $(9N^2,n_w)$, where $n_w \simeq 9N^2/2$.
Despite having $9N^2$ non-zero elements, the total number of elements in $\bm{B}_w$ is much larger, highlighting the sparsity of the eigenvectors.

For third-order force constants, the permutation symmetry rule for $(i\alpha,j\beta,k\gamma)$ is specified by Eq. (\ref{symfc:eq-permutation-fc3}), which comprises five distinct equations.
This symmetry rule can be represented in matrix form as $\bm{W}_{i\alpha,j\beta,k\gamma}^\top \bm{\theta}=0$, and the submatrix of $\bm{W}_{i\alpha,j\beta,k\gamma}$, constructed by extracting only the non-zero rows, is given by
\begin{equation}
\bm{W}_{i\alpha,j\beta,k\gamma} = \frac{1}{\sqrt{2}}
\begin{bmatrix}
1  & 1 & 1 & 1 & 1 \\
-1 & 0 & 0 & 0 & 0 \\
0 & -1 & 0 & 0 & 0 \\
0 & 0 & -1 & 0 & 0 \\
0 & 0 & 0 & -1 & 0 \\
0 & 0 & 0 & 0 & -1 \\
\end{bmatrix}
.
\end{equation}
Therefore, the column vector $\bm{w}_{i\alpha,j\beta,k\gamma}^\perp$, which is orthonormal to all column vectors of $\bm{W}_{i\alpha,j\beta,k\gamma}$, can be considered as an eigenvector of $\bm{P}_w$. 
It is given by
\begin{equation}
\bm{w}_{i\alpha,j\beta,k\gamma}^\perp = \frac{1}{\sqrt{6}}[1\:1\:1\:1\:1\:1]^\top,
\end{equation}
which is represented in the submatrix form by omitting zero elements.
When $i\alpha=j\beta$ and $i\alpha \neq k\gamma$, the permutation symmetry rule for $(i\alpha,i\alpha,k\gamma)$ is given by Eq. (\ref{symfc:eq-permutation-fc3-2}).
The two equations in Eq. (\ref{symfc:eq-permutation-fc3-2}) are also represented in matrix form as $\bm{W}_{i\alpha,i\alpha,k\gamma}^\top \bm{\theta}=0$, where the submatrix of $\bm{W}_{i\alpha,i\alpha,k\gamma}$ comprising only non-zero rows is expressed as
\begin{equation}
\bm{W}_{i\alpha,i\alpha,k\gamma} = \frac{1}{\sqrt{2}}
\begin{bmatrix}
1  & 1 \\
-1 & 0 \\
0 & -1 \\
\end{bmatrix}
.
\end{equation}
Consequently, the vector orthonormal to all column vectors of $\bm{W}_{i\alpha,j\beta,k\gamma}$, denoted by $\bm{w}_{i\alpha,i\alpha,k\gamma}^\perp$, is given by
\begin{equation}
\bm{w}_{i\alpha,i\alpha,k\gamma}^\perp = \frac{1}{\sqrt{3}}[1\:1\:1]^\top.
\end{equation}
If $i\alpha=j\beta$ and $i\alpha=k\gamma$, there are no permutation symmetry rules applicable. 
In this case, the vector $\bm{w}_{i\alpha,i\alpha,i\alpha}^\perp$ is defined as the vector with a single non-zero element equal to one for the column corresponding to $(i,\alpha,i,\alpha,i,\alpha)$.
Thus, the eigenvectors $\bm{B}_w$ are written as
\begin{equation}
\bm{B}_w = [\cdots, \bm{w}_{i\alpha,j\beta,k\gamma}^\perp, \cdots],
\end{equation}
where the vectors $\bm{w}_{i\alpha,j\beta,k\gamma}^\perp$ are mutually orthonormal.
The matrix has dimensions $(27N^3,n_w)$, where $n_w \simeq 27N^3/6$. 
Although the number of non-zero elements in $\bm{B}_w$ is $27N^3$,  it is significantly smaller than the total number of elements in $\bm{B}_w$.

\subsection{Effective use of lattice translation group}
Here, we introduce the projection matrix corresponding to the identity irreducible representation of the lattice translation group, which is a normal subgroup of the space group.
While the inclusion of the projection matrix for the lattice translation group is not strictly necessary for the comprehensive description of the full projection matrix, it is advantageous for the compression of the projection matrices that constitute components of the full projection matrix.
The projection matrix for the lattice translation group can be derived using a methodology analogous to that employed for the projection matrix associated with permutation symmetry rules.

The force constants are invariant under operations from the lattice translation group $\mathcal{T} = \{t_1,t_2,\cdots\}$.
Here, representative atoms with respect to lattice translations are denoted by indices $\{i'\}$, and the translation of atom $i$ is represented as $t_n(i)$.
For the second-order force constants involving a representative atom $i'$, the invariance under a lattice translation $t_n$ is expressed by
\begin{equation}
\Theta_{i'\alpha,j\beta} - \Theta_{t_n(i')\alpha,t_n(j)\beta} = 0,
\end{equation}
where $t_n$ is a non-identity translation operation.
Consequently, for each pair $(i',\alpha,j,\beta)$, there are $|\mathcal{T}|-1$ invariance equations, which can be represented in matrix form by $\bm{T}_{i'\alpha,j\beta}$ such that $\bm{T}_{i'\alpha,j\beta}^\top \bm{\theta} = 0$.
The submatrix of $\bm{T}_{i'\alpha,j\beta}$ composed only of non-zero rows is expressed as
\begin{equation}
\label{symfc:eq-T-submatrix}
\bm{T}_{i',\alpha,j\beta} = \frac{1}{\sqrt{2}}
\begin{bmatrix}
1  & 1 & 1 & 1 &  \\
-1 & 0 & 0 & 0 &  \\
0 & -1 & 0 & 0 & \cdots \\
0 & 0 & -1 & 0 &  \\
0 & 0 & 0 & -1 &  \\
 & \vdots & &  & \ddots \\
\end{bmatrix},
\end{equation}
where the size of this submatrix is $(|\mathcal{T}|,|\mathcal{T}|-1)$.
The vector orthonormal to all column vectors of $\bm{T}_{i'\alpha,j\beta}$ can be an eigenvector of $\bm{P}_{\rm LT}$. 
This eigenvector is represented in the submatrix form as
\begin{equation}
\label{symfc:eq-t-perp}
\bm{t}_{i'\alpha,j\beta}^\perp = \frac{1}{\sqrt{|\mathcal{T}|}} [1\:1\:1\:\cdots]^\top.
\end{equation}
The eigenvectors of $\bm{P}_{\rm LT}$, denoted by $\bm{B}_{\rm LT}$, are then given by
\begin{equation}
\bm{B}_{\rm LT} = [\cdots, \bm{t}_{i'\alpha,j\beta}^\perp, \cdots],
\end{equation}
where the set of vectors $\{\bm{t}_{i'\alpha,j\beta}^\perp\}$ is orthonormal.
The size of $\bm{B}_{\rm LT}$ is $(9N^2,9N^2/|\mathcal{T}|)$, and it contains $9N^2$ non-zero elements.

The third-order force constants for triplet atoms including a representative atom $i'$ are also invariant under the operations of the lattice translation group $\mathcal{T}$. 
This invariance is expressed as
\begin{equation}
\Theta_{i'\alpha,j\beta,k\gamma} - \Theta_{t_n(i')\alpha,t_n(j)\beta,t_n(k)\gamma} = 0,
\end{equation}
where $t_n$ is a translation operation that is not the identity.
Similarly to the case of second-order force constants, there are $|\mathcal{T}|-1$ equations for each triplet of indices $(i',\alpha,j,\beta,k,\gamma)$, which can be represented in matrix form as $\bm{T}_{i'\alpha,j\beta,k\gamma}^\top \bm{\theta}^{(\rm FC3)} = 0$.
The submatrix forms of $\bm{T}_{i'\alpha,j\beta,k\gamma}$ and $\bm{t}_{i'\alpha,j\beta,k\gamma}^\perp$ are obtained by extending the indices from ($i'\alpha,j\beta$) to ($i'\alpha,j\beta,k\gamma$) in Eqs. (\ref{symfc:eq-T-submatrix}) and (\ref{symfc:eq-t-perp}).
Therefore, the eigenvectors of $\bm{P}_{\rm LT}$ for the third-order force constants can be expressed as
\begin{equation}
\bm{B}_{\rm LT} = [\cdots, \bm{t}_{i'\alpha,j\beta,k\gamma}^\perp, \cdots],
\end{equation}
where the set of vectors $\{\bm{t}_{i'\alpha,j\beta,k\gamma}^\perp\}$ is orthonormal.
The size of $\bm{B}_{\rm LT}$ is $(27N^3,27N^3/|\mathcal{T}|)$, and the number of non-zero elements is $27N^3$, which is considerably smaller than the total number of elements in $\bm{B}_{\rm LT}$.
Thus, $\bm{B}_{\rm LT}$ remains a sparse matrix.

The projection matrix $\bm{P}_{\rm LT}$ for the lattice translation group is given by its eigenvectors as $\bm{P}_{\rm LT} = \bm{B}_{\rm LT} \bm{B}_{\rm LT}^\top$.
The projection matrix for the lattice translation group satisfies the following relationship with the projection matrix $\bm{P}_{\rm spg}$ for the space group:
\begin{equation}
\label{symfc:eq-lt-spg}
\bm{P}_{\rm LT} \bm{P}_{\rm spg} = \bm{P}_{\rm spg} \bm{P}_{\rm LT} = \bm{P}_{\rm spg}.
\end{equation}
This equation holds because we have $t_n \mathcal{G} = \mathcal{G}$ for any translational operation $t_n \in \mathcal{T}$ that is also included in the group $\mathcal{G}$.
Therefore, the projection matrix for the lattice translation group leaves $\bm{P}_{\rm spg}$ invariant.

\subsection{Properties of projection matrices}
The commutation relations between different projection matrices play a crucial role in the efficient construction of force-constant basis sets. 
These relations include
\begin{equation}
[\bm{P}_{\rm spg}, \bm{P}_v] = 0, \: [\bm{P}_{\rm spg}, \bm{P}_w] = 0,\: [\bm{P}_v, \bm{P}_w] \neq 0.
\end{equation}
Additionally, the projection matrices for the lattice translation group $\bm{P}_{\rm LT}$ exhibit useful commutation properties expressed as
\begin{equation}
[\bm{P}_{\rm LT},\bm{P}_v]=0,\: [\bm{P}_{\rm LT}, \bm{P}_w] = 0,
\end{equation}
in conjunction with the relationship of Eq. (\ref{symfc:eq-lt-spg}).
In addition, all these projection matrices are idempotent and symmetric, which are represented by
\begin{equation}
\bm{P}_{\rm spg}^2 = \bm{P}_{\rm spg}, \:
\bm{P}_v^2 = \bm{P}_v, \:
\bm{P}_w^2 = \bm{P}_w, \:
\bm{P}_{\rm LT}^2 = \bm{P}_{\rm LT}
\end{equation}
and 
\begin{equation}
\bm{P}_{\rm spg}^\top = \bm{P}_{\rm spg}, \:
\bm{P}_v^\top = \bm{P}_v, \:
\bm{P}_w^\top = \bm{P}_w, \:
\bm{P}_{\rm LT}^\top = \bm{P}_{\rm LT},
\end{equation}
respectively.
% These properties ensure that the matrices are orthogonal projections onto their respective subspaces.

\subsection{Eigenvalue problems for projection matrices}
\label{symfc:sec-eigenvalue-problem}

Solving eigenvalue problems for large and sparse projection matrices is a fundamental element of the current procedure.
As demonstrated in Sec. \ref{symfc:sec-procedure-basis-set}, the efficiency of the solver is crucial in determining the design of the procedure. 
When addressing eigenvalue problems for sparse projection matrices, exploiting the underlying block-diagonal structures can be particularly advantageous.

We consider an orthogonal projection matrix expressed in block-diagonal form, which is transformed by row and column permutations.
The block-diagonal form of the orthogonal projection matrix is given by
\begin{equation}
\bm{M}^\top \bm{P} \bm{M} = 
\begin{bmatrix}
\bm{I} & \bm{0} & \bm{0} \\
\bm{0} & \bm{J} & \bm{0} \\
\bm{0} & \bm{0} & \bm{K} \\
\end{bmatrix}
,
\end{equation}
where $\bm{M}$ denotes the permutation matrix for columns.
The submatrices $\bm{I}$, $\bm{J}$, and $\bm{K}$ are all orthogonal projection matrices, ensuring that $\bm{P}$ is also an orthogonal projection matrix.
When solving eigenvalue problems for these submatrices, and obtaining the eigenvectors of $\bm{I}$, $\bm{J}$, and $\bm{K}$ as
\begin{equation}
\bm{I}\bm{b_I} = \bm{b_I},\:\bm{J}\bm{b_J} = \bm{b_J}, \: \bm{K}\bm{b_K} = \bm{b_K},
\end{equation}
the vectors 
\begin{equation}
\bm{b} = 
\begin{bmatrix}
\bm{b_I} \\
\bm{0}   \\
\bm{0}   \\
\end{bmatrix},
\begin{bmatrix}
\bm{0} \\
\bm{b_J}   \\
\bm{0}   \\
\end{bmatrix},
\begin{bmatrix}
\bm{0} \\
\bm{0}   \\
\bm{b_K}   \\
\end{bmatrix}
\end{equation}
can be adopted as eigenvectors of $\bm{M}^\top\bm{P}\bm{M}$.
The set of eigenvectors $\{\bm{b}\}$ is orthonormal, and the trace of $\bm{M}^\top\bm{P}\bm{M}$ is given by
\begin{equation}
\operatorname{tr}(\bm{M}^\top\bm{P}\bm{M}) = 
\operatorname{tr}(\bm{I}) + \operatorname{tr}(\bm{J}) + \operatorname{tr}(\bm{K}),
\end{equation}
which equals the number of eigenvectors of $\bm{M}^\top\bm{P}\bm{M}$.
Consequently, the set of eigenvectors $\{\bm{b}\}$ constructed from $\{\bm{b_I}\}$, $\{\bm{b_J}\}$, and $\{\bm{b_K}\}$ forms a complete basis.
Finally, the eigenvectors of $\bm{P}$ can be computed as $\bm{M} \bm{b}$.
Thus, the eigenvectors of the original projection matrix $\bm{P}$ can be determined by solving the eigenvalue problems for the smaller matrices.

We use a graph-based approach to determine the permutation matrix $\bm{M}$ and reveal the block-diagonal structure of $\bm{P}$. 
We represent the orthogonal projection matrix $\bm{P}$ as an undirected graph, where each row and column corresponds to a vertex. 
Vertices
$s$ and $t$ are adjacent if the element in $\bm{P}$ at row $s$ and column $t$ is non-zero.
We then apply an efficient graph algorithm to identify the connected components \cite{10.5555/1408038} of this graph. 
The algorithm groups the rows and columns of $\bm{P}$ into connected components, which correspond to block structures in the matrix. 
Finally, we construct the permutation matrix $\bm{M}$ so that columns within the same connected component are placed adjacent to each other in the reordered matrix $\bm{M}^\top\bm{P}\bm{M}$.

\subsection{Procedure to calculate eigenvectors of full projection matrix}
\label{symfc:sec-procedure-basis-set}

As described in Sec. \ref{symfc:sec-non-commutative-matrices}, the eigenvectors of the approximated projection matrix $\tilde{\bm{P}} = \bm{P}_{\rm spg} \bm{P}_{w} \bm{P}_{v} \bm{P}_{w}$ corresponding to eigenvalues equal to one constitute a complete basis for the full projection matrix $\bm{P}$.
The computational steps for determining the eigenvectors of $\bm{P}$ are outlined as follows.

(1) The projection matrix $\bm{P}_{w}$ is compressed using $\bm{B}_{\rm LT}$.
The compressed projection matrix is given by
\begin{equation}
\bm{P}_{w \downarrow {\rm LT}} = \bm{B}_{\rm LT}^\top \bm{B}_{w} \bm{B}_{w}^\top \bm{B}_{\rm LT}.
\end{equation}
Thus, the product $\bm{P}_{\rm spg} \bm{P}_{w}$ can be rewritten as
\begin{eqnarray}
\bm{P}_{\rm spg} \bm{P}_{w}
&=& \bm{P}_{\rm spg} \bm{P}_{\rm LT} \bm{P}_{w} \bm{P}_{\rm LT} \nonumber \\
&=& \bm{P}_{\rm spg} \bm{B}_{\rm LT} \bm{B}_{\rm LT}^\top \bm{B}_{w} \bm{B}_{w}^\top \bm{B}_{\rm LT} \bm{B}_{\rm LT}^\top \\
&=& \bm{P}_{\rm spg} \bm{B}_{\rm LT} \bm{P}_{w \downarrow {\rm LT}} \bm{B}_{\rm LT}^\top, \nonumber
\end{eqnarray}
where the properties of $\bm{P}_{\rm LT}$ and Eq. (\ref{symfc:matrix-comp-definition2}) are utilized in the derivation.

(2) The eigenvalue problem for $\bm{P}_{w \downarrow {\rm LT}}$ is solved. 
The compressed projection matrix $\bm{P}_{w \downarrow {\rm LT}}$ is decomposed as
\begin{equation}
\bm{P}_{w \downarrow {\rm LT}} = \bm{B}_{w \downarrow {\rm LT}} \bm{B}_{w \downarrow {\rm LT}}^\top.
\end{equation}
Using this decomposition, the product of projection matrices $\bm{P}_{\rm LT} \bm{P}_{w} \bm{P}_{\rm LT}$ is decomposed as
\begin{eqnarray}
\bm{P}_{\rm LT} \bm{P}_{w} \bm{P}_{\rm LT}
&=& \bm{B}_{\rm LT} \bm{B}_{w \downarrow {\rm LT}} \bm{B}_{w \downarrow {\rm LT}}^\top \bm{B}_{\rm LT}^\top \nonumber \\
&=& \bm{B}_{w \cap {\rm LT}} \bm{B}_{w \cap {\rm LT}}^\top,
\end{eqnarray}
where 
\begin{equation}
\bm{B}_{w \cap {\rm LT}} = \bm{B}_{\rm LT} \bm{B}_{w \downarrow {\rm LT}}.
\end{equation}
The set of vectors $\bm{B}_{w \cap {\rm LT}}$ spans the intersection of the subspaces spanned by $\bm{B}_{w}$ and $\bm{B}_{\rm LT}$.

(3) The matrix $\bm{P}_{\rm spg}$ is compressed using $\bm{B}_{w \cap {\rm LT}}$ as follows:
\begin{equation}
\bm{P}_{{\rm spg} \downarrow (w \cap {\rm LT})} = \bm{B}_{w \cap {\rm LT}}^\top \bm{P}_{\rm spg} \bm{B}_{w \cap {\rm LT}}.
\end{equation}
Using this compressed projection matrix, the product $\bm{P}_{\rm spg} \bm{P}_{w}$ is expressed as
\begin{eqnarray}
\bm{P}_{\rm spg} \bm{P}_{w}
&=& \bm{P}_{\rm LT} \bm{P}_{w} \bm{P}_{\rm LT} \bm{P}_{\rm spg} \bm{P}_{\rm LT} \bm{P}_{w} \bm{P}_{\rm LT} \nonumber \\
&=& \bm{B}_{w \cap {\rm LT}} \bm{B}_{w \cap {\rm LT}}^\top \bm{P}_{\rm spg} \bm{B}_{w \cap {\rm LT}} \bm{B}_{w \cap {\rm LT}}^\top \\
&=& \bm{B}_{w \cap {\rm LT}} \bm{P}_{{\rm spg} \downarrow (w \cap {\rm LT})} \bm{B}_{w \cap {\rm LT}}^\top \nonumber.
\end{eqnarray}

(4) The eigenvalue problem for $\bm{P}_{{\rm spg} \downarrow (w \cap {\rm LT})}$ is solved, and it is decomposed as
\begin{equation}
\bm{P}_{{\rm spg} \downarrow (w \cap {\rm LT})} = \bm{B}_{{\rm spg} \downarrow (w \cap {\rm LT})} \bm{B}_{{\rm spg} \downarrow (w \cap {\rm LT})}^\top.
\end{equation}
Consequently, the product $\bm{P}_{\rm spg} \bm{P}_{w}$ is expressed as
\begin{eqnarray}
\bm{P}_{\rm spg} \bm{P}_{w}
&=& \bm{B}_{w \cap {\rm LT}} \bm{B}_{{\rm spg} \downarrow (w \cap {\rm LT})} \bm{B}_{{\rm spg} \downarrow (w \cap {\rm LT})}^\top \bm{B}_{w \cap {\rm LT}}^\top \nonumber \\
&=& \bm{B}_{{\rm spg} \cap w} \bm{B}_{{\rm spg} \cap w}^\top,
\end{eqnarray}
where 
\begin{equation}
\bm{B}_{{\rm spg} \cap w} = \bm{B}_{w \cap {\rm LT}} \bm{B}_{{\rm spg} \downarrow (w \cap {\rm LT})}.
\end{equation}

(5) The projection matrix $\bm{P}_{v}$ is compressed using $\bm{B}_{{\rm spg} \cap w}$ as
\begin{equation}
\bm{P}_{v\downarrow({\rm spg} \cap w)} = \bm{B}_{{\rm spg} \cap w}^\top \bm{P}_{v} \bm{B}_{{\rm spg} \cap w}.
\end{equation}
Using this compressed form of $\bm{P}_{v}$, the approximated projection matrix $\tilde{\bm{P}}$ is represented by
\begin{eqnarray}
\label{symfc:eq-procedure2}
\tilde{\bm{P}} 
&=& \bm{P}_{\rm spg} \bm{P}_{w} \bm{P}_{v} \bm{P}_{w} \nonumber \\
&=& \bm{P}_{\rm spg} \bm{P}_{w} \bm{P}_{v} \bm{P}_{\rm spg} \bm{P}_{w} \nonumber \\
&=& \bm{B}_{{\rm spg} \cap w} \bm{B}_{{\rm spg} \cap w}^\top \bm{P}_{v} \bm{B}_{{\rm spg} \cap w} \bm{B}_{{\rm spg} \cap w}^\top \nonumber \\
&=& \bm{B}_{{\rm spg} \cap w} \bm{P}_{v\downarrow({\rm spg} \cap w)} \bm{B}_{{\rm spg} \cap w}^\top.
\end{eqnarray}

(6) The eigenvalue problem for $\bm{P}_{v\downarrow({\rm spg} \cap w)}$ is solved. 
By eliminating the eigenvectors associated with eigenvalues less than one, the full projection matrix $\bm{P}$ is obtained as
\begin{eqnarray}
\label{symfc:eq-procedure-eig-sum}
\bm{P}
&=& \bm{B}_{{\rm spg} \cap w} \bm{B}_{v\downarrow({\rm spg} \cap w)} \bm{B}_{v\downarrow({\rm spg} \cap w)}^\top \bm{B}_{{\rm spg} \cap w}^\top \nonumber \\
&=& \bm{B} \bm{B}^\top,
\end{eqnarray}
where the basis set for the force constants is given by
\begin{equation}
\label{symfc:eqn:full-basis}
\bm{B} = \bm{B}_{{\rm spg} \cap w} \bm{B}_{v\downarrow({\rm spg} \cap w)}.
\end{equation}

\subsection{Practical approximations}
The computational bottleneck in the current procedure arises from solving the eigenvalue problem specified by Eq. (\ref{symfc:eq-procedure-eig-sum}), particularly as the size of the basis set increases. 
Empirical observations indicate that the eigenvalue problem becomes computationally intensive when the dimension of the square projection matrix $\bm{P}_{v\downarrow({\rm spg} \cap w)}$ exceeds 20000.
To alleviate this issue, an approximation can be employed to reduce the basis set by assigning zero values to a subset of force constant elements that are expected to be near zero.
Specifically, when third-order force constants with indices $\mathbb{S}$ are set to zero, the projection matrix designed to eliminate these zero elements is given by
\begin{equation}
\bm{P}_{\rm NZ} = \bm{I} - \sum_{(i\alpha,j\beta,k\gamma) \in \mathbb{S}} \bm{\eta}_{i\alpha,j\beta,k\gamma} \bm{\eta}_{i\alpha,j\beta,k\gamma} ^\top,
\end{equation}
where $\bm{\eta}_{i\alpha,j\beta,k\gamma}$ represents a vector with a single non-zero element equal to one, corresponding to each element in $\mathbb{S}$.
Consequently, the projection matrix $\bm{P}_{\rm NZ}$ is a diagonal matrix with zero entries for indices in $\mathbb{S}$ and one entries for all other indices.
Since $\bm{P}_{\rm NZ}$ and a projection matrix $\bm{P}_1$ are generally non-commutative, i.e., $[\bm{P}_{\rm NZ}, \bm{P}_1] \neq 0$, a basis set spanning the intersection of vector subspaces determined by the projection matrices can be obtained by solving the eigenvalue problem for $\bm{P}_{\rm NZ} \bm{P}_1 \bm{P}_{\rm NZ}$ and discarding eigenvectors with eigenvalues less than one, as shown in Sec. \ref{symfc:sec-non-commutative-matrices}.

In practical terms, force constant elements are set to zero by eliminating the basis vectors corresponding to $\mathbb{S}$ from $\bm{B}_w$ and $\bm{B}_{\rm LT}$.
Specifically, when $\Theta_{i\alpha,j\beta,k\gamma}$ is set to zero, the column basis vectors in $\bm{B}_w$ and $\bm{B}_{\rm LT}$ that contain non-zero elements associated with the row $(i\alpha,j\beta,k\gamma)$ are removed.
This means that projection matrices are applied to $\bm{P}_w$ and $\bm{P}_{\rm LT}$ as follows:
\begin{equation}
\bm{P}_{w \cap {\rm NZ}} = \left[ \bm{I} - \sum_i \bm{b}_{w,i} \bm{b}_{w,i}^\top \right] \bm{P}_w
\end{equation}
and
\begin{equation}
\bm{P}_{{\rm LT} \cap {\rm NZ}} = \left[\bm{I} - \sum_i \bm{b}_{{\rm LT},i} \bm{b}_{{\rm LT},i}^\top \right] \bm{P}_{\rm LT},
\end{equation}
where $\bm{b}_{w,i}$ and $\bm{b}_{{\rm LT},i}$ represent the basis vectors removed from $\bm{B}_w$ and $\bm{B}_{\rm LT}$, respectively.
These projection matrices can be utilized by substituting $\bm{P}_w$ and $\bm{P}_{\rm LT}$ in the formulation of basis set construction.

A systematic approach to introducing zero elements involves employing a cutoff distance. 
The force constant $\Theta_{i\alpha,j\beta,k\gamma}$ is set to zero if the distance between any pair of atoms $i$, $j$, and $k$ exceeds the cutoff distance. 
In this study, the distance between two atoms within a given periodic supercell is defined as the minimum distance among those calculated using both the atoms in the supercell and the replica atoms located in adjacent supercells.

\section{Force constant estimation}
\label{symfc:sec-force-constant-estimation}
\subsection{Force-constant basis set expansion}

The forces acting on atoms $\{f_{i\alpha}^{(s)}\}$ in a supercell structure $s$ corresponding to atomic displacements $\{u_{i\alpha}^{(s)}\}$ is written as
\begin{equation}
f_{i\alpha}^{(s)} = 
- \sum_{j\beta} \Theta_{i\alpha,j\beta} u_{j\beta}^{(s)}
- \frac{1}{2} \sum_{j\beta} \sum_{k\gamma} \Theta_{i\alpha,j\beta,k\gamma} u_{j\beta}^{(s)} u_{k\gamma}^{(s)}
+ \cdots.
\end{equation}
The force constants are then expanded in basis sets given by Eq. (\ref{symfc:eqn:full-basis}) as 
\begin{equation}
\Theta_{i\alpha,j\beta} = \sum_b c_b^{\rm (FC2)} B_{i\alpha,j\beta,b}^{\rm (FC2)}
\end{equation}
and
\begin{equation}
\Theta_{i\alpha,j\beta,k\gamma} = \sum_b c_b^{\rm (FC3)} B_{i\alpha,j\beta,k\gamma,b}^{\rm (FC3)},
\end{equation}
where $c_b$ denotes the expansion coefficient for $b$-th basis.
Using these basis expansions, the forces in structure $s$ are expressed as
\begin{eqnarray}
f_{i\alpha}^{(s)} = 
&-& \sum_b c_b^{\rm (FC2)} \sum_{j\beta} u_{j\beta}^{(s)} B_{i\alpha,j\beta,b}^{\rm (FC2)}
\nonumber \\
&-& \frac{1}{2} \sum_b c_b^{\rm (FC3)} \sum_{j\beta} \sum_{k\gamma} u_{j\beta}^{(s)} u_{k\gamma}^{(s)} B_{i\alpha,j\beta,k\gamma,b}^{\rm (FC3)}
\nonumber \\
&+& \cdots,
\end{eqnarray}
%\begin{eqnarray}
%f_{i\alpha}^{(s)} = 
%&-& \sum_b c_b^{\rm (FC2)} \sum_{j\beta} \sum_c u_{j\beta}^{(s)} C_{i\alpha,j\beta,c}^{\rm (FC2)} E_{c,b}^{\rm (FC2)} 
%\nonumber \\
%&-& \frac{1}{2} \sum_b c_b^{\rm (FC3)} \sum_{j\beta} \sum_{k\gamma} \sum_c u_{j\beta}^{(s)} u_{k\gamma}^{(s)} C_{i\alpha,j\beta,k\gamma,c}^{\rm (FC3)} E_{c,b}^{\rm (FC3)} 
%\nonumber \\
%&+& \cdots.
%\end{eqnarray}
and they are rewritten as
\begin{eqnarray}
f_{i\alpha}^{(s)} = 
\sum_b c_b^{\rm (FC2)} x_{i\alpha,b}^{(s,{\rm FC2})}
+ \sum_b c_b^{\rm (FC3)} x_{i\alpha,b}^{(s,{\rm FC3})}
+ \cdots,
\end{eqnarray}
where
\begin{eqnarray}
x_{i\alpha,b}^{(s,{\rm FC2})} = 
- \sum_{j\beta} u_{j\beta}^{(s)} B_{i\alpha,j\beta,b}^{\rm (FC2)}
\end{eqnarray}
and
\begin{eqnarray}
x_{i\alpha,b}^{(s,{\rm FC3})} = 
- \frac{1}{2} \sum_{j\beta} \sum_{k\gamma} u_{j\beta}^{(s)} u_{k\gamma}^{(s)} B_{i\alpha,j\beta,k\gamma,b}^{\rm (FC3)}.
\end{eqnarray}
Thus, when modeling forces up to third-order force constants, the forces in structure $s$ are represented in matrix form as given by
\begin{equation}
\bm{f}^{(s)} = \bm{X}^{(s)} \bm{c},
\end{equation}
where
\begin{equation}
  \bm{f}^{(s)} = 
  \begin{bmatrix}
  \vdots \\
  f_{i\alpha}^{(s)} \\
  \vdots \\
  \end{bmatrix},
\end{equation}

\begin{eqnarray}
  \bm{X}^{(s)} &=&
  \begin{bmatrix}
  \bm{X}^{(s, \rm FC2)} & \bm{X}^{(s, \rm FC3)} \\
  \end{bmatrix} 
  \nonumber \\
  &=&
  \begin{bmatrix}
  \cdots & \vdots & \cdots & \vdots & \cdots \\
  \cdots & x_{i\alpha,b}^{(s,{\rm FC2})} &\cdots& x_{i\alpha,b}^{(s,{\rm FC3})} & \cdots\\
  \cdots & \vdots & \cdots & \vdots & \cdots \\
  \end{bmatrix}, 
\end{eqnarray}
and
\begin{equation}
  \bm{c} = 
  \begin{bmatrix}
  \bm{c}^{(\rm FC2)} \\
  \bm{c}^{(\rm FC3)} \\
  \end{bmatrix}
  =
  \begin{bmatrix}
  \vdots \\
  c_{b}^{\rm (FC2)} \\
  \vdots \\
  c_{b}^{\rm (FC3)} \\
  \vdots \\
  \end{bmatrix}.
\end{equation}

\subsection{Force constant estimation}
The expansion coefficients \(\bm{c}\) are determined from the forces acting on atoms in a training displacement-force dataset consisting of \(N_s\) structures using standard linear regression.
The training dataset is expressed in matrix form as
\begin{equation}
  \bm{X} = 
  \begin{bmatrix}
  \bm{X}^{(1)} \\
  \bm{X}^{(2)} \\
  \bm{X}^{(3)} \\
  \vdots \\
  \bm{X}^{(N_s)}
  \end{bmatrix}, 
  \:\:
  \bm{y} = 
  \begin{bmatrix}
  \bm{f}^{(1)} \\
  \bm{f}^{(2)} \\
  \bm{f}^{(3)} \\
  \vdots \\
  \bm{f}^{(N_s)}
  \end{bmatrix},
\end{equation}
and the normal equations to find the least squares solution are given by 
\begin{equation}
\bm{X}^\top \bm{X} \bm{c} = \bm{X}^\top\bm{y}.
\end{equation}
Once the coefficients $\bm{c}$ are estimated, the force constants can then be recovered from these coefficients using
\begin{eqnarray}
\bm{\theta}^{(\rm FC2)} = \bm{B}^{(\rm FC2)} \bm{c}^{(\rm FC2)} \nonumber \\
\bm{\theta}^{(\rm FC3)} = \bm{B}^{(\rm FC3)} \bm{c}^{(\rm FC3)}.
\end{eqnarray}

Here, we consider solving the normal equations directly. However, computing the entire matrix $\bm{X}$ can be resource-intensive, particularly with large displacement-force datasets and extensive basis sets.
To optimize memory usage and efficiently estimate the expansion coefficients, we split the computations of $\bm{X}^\top \bm{X}$ and execute them sequentially. 
This approach is based on the recursive method for linear regression \cite{strang2019linear}.
When the dataset is partitioned into $n_q$ subsets, $\bm{X}^\top \bm{X}$ and $\bm{X}^\top\bm{y}$ can be calculated as
\begin{equation}
\bm{X}^\top \bm{X} = \sum_{q=1}^{n_q} \bm{X}_q^\top \bm{X}_q, 
\:\:\:
\bm{X}^\top \bm{y} = \sum_{q=1}^{n_q} \bm{X}_q^\top \bm{y}_q,
\end{equation}
where $\bm{X}_q$ and $\bm{y}_q$ represent the $q$-th subsets of $\bm{X}$ and $\bm{y}$, respectively.
This approach eliminates the need to store the entire matrix $\bm{X}$ in memory.
The coefficients are then determined by solving the normal equations using fast linear algebra libraries.
A more detailed implementation for efficiently computing $\bm{X}^\top \bm{X}$ and $\bm{X}^\top\bm{y}$ is provided in Appendix \ref{symfc:sec-appendix-fc-estimation}.

Note that the methodology for estimating higher-order force constants can be similarly formalized. 
The symfc code is capable of performing force constant estimation up to the fourth order.

\section{Results and discussion}
\label{symfc:sec-results}
\subsection{Basis set construction}
\subsubsection{Complete basis set}

\begin{figure*}[tbp]
\includegraphics[clip,width=\linewidth]{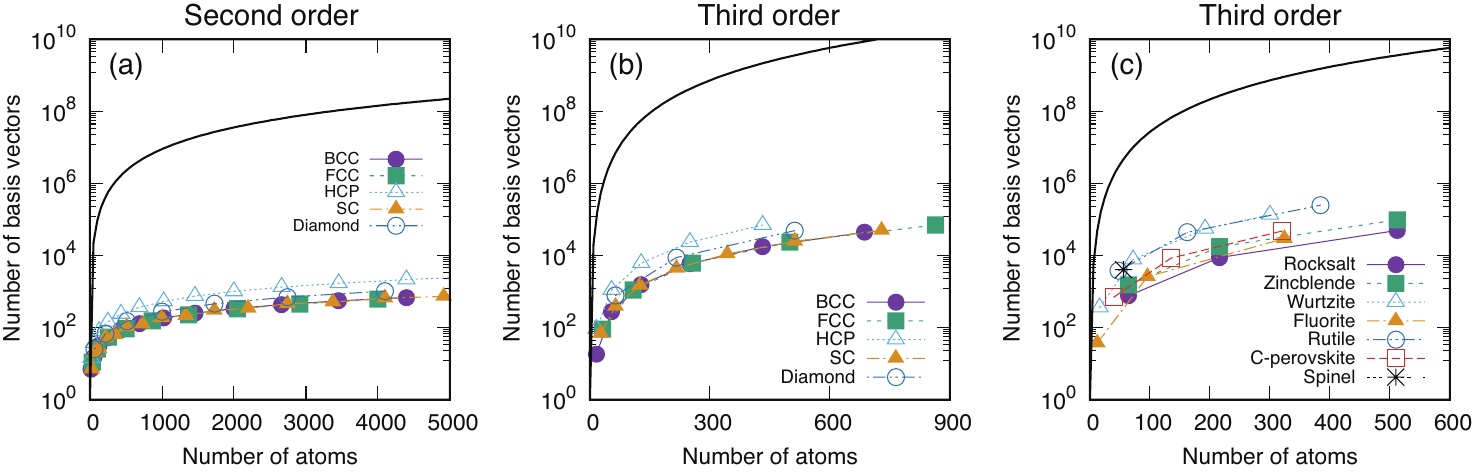}
\caption{
Number of basis vectors required to completely represent force constants in different crystal structures. 
(a) For second-order force constants in elemental crystal structures: body-centered cubic (bcc), face-centered cubic (fcc), hexagonal close-packed (hcp), simple cubic (sc), and diamond-type structures. 
(b) For third-order force constants in elemental crystal structures. 
(c) For third-order force constants in ionic crystal structures, including rocksalt, zincblende, wurtzite, fluorite, rutile, cubic perovskite (C-perovskite), and spinel structures. 
The black solid lines indicate the theoretical values of $9N^2$ for second-order force constants and $27N^3$ for third-order force constants.
}
\label{symfc:Fig-num-basis}
\end{figure*}

For various structures and supercell sizes, orthonormal basis sets for the representation of second-order and third-order supercell force constants are determined using the current projector-based approach implemented in the symfc code.
Given a unit cell structure and a specified supercell size, a basis set for the force constants is derived by applying the procedure shown in Sec. \ref{symfc:sec-procedure-basis-set}.
This basis set ensures that the supercell force constants completely satisfy the permutational symmetry requirements, sum rules, and the symmetric properties of the supercell.

Figure \ref{symfc:Fig-num-basis} (a) illustrates the size of the basis set for the second-order force constants of various crystal structures for elemental systems. 
The total number of second-order force constant elements is 9$N^2$, but this number is significantly reduced by applying constraints on the force constants. 
For second-order force constants, it is possible to derive complete basis sets for supercells containing up to 10000 atoms and approximately $10^9$ force constant elements, while still maintaining acceptable computational efficiency.

The total number of third-order force-constant elements is 27$N^3$, which can become quite substantial when evaluating force constants for supercells of typical size. 
Consequently, the reduction of third-order force constants is significantly more computationally demanding compared to that of second-order force constants for a given $N$.
The current projector-based approach, however, efficiently reduces the number of force constants and constructs orthonormal basis sets.

Figures \ref{symfc:Fig-num-basis} (b) and (c) display the number of basis vectors required for third-order force constants in elemental and ionic structures, respectively. 
In these figures, the cutoff distance is not applied, and no force-constant elements are constrained to be zero.
The computational bottleneck in the current procedure arises from solving the eigenvalue problem given by Eq. (\ref{symfc:eq-procedure-eig-sum}), particularly when the basis set size becomes large. 
Thanks to the small primitive cells and high symmetry of elemental structures, the current procedure can be applied to supercells with up to 900 atoms using a standard workstation. 

In contrast, for ionic structures, the number of atoms in applicable supercells is reduced due to decreased lattice translations. 
For the well-known structures shown in Fig. \ref{symfc:Fig-num-basis} (c), the procedure generally performs well for supercells containing up to 300 atoms. 
Additional details on the computational costs associated with constructing complete basis sets are provided in Appendix \ref{symfc:sec:comptational-efficiency}.

\subsubsection{Application of cutoff distance}

\begin{figure}[tbp]
\includegraphics[clip,width=0.8\linewidth]{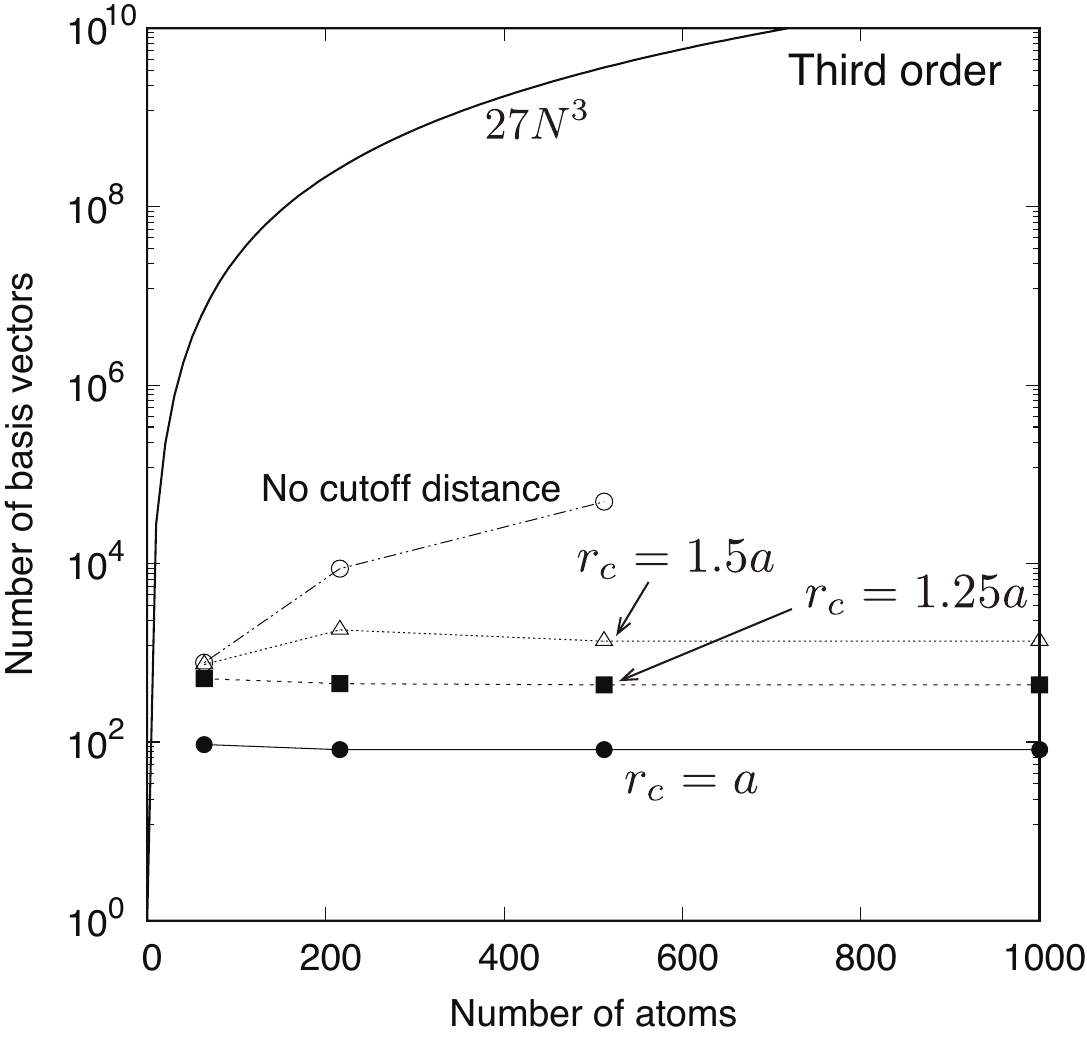}
\caption{
Number of basis vectors to represent third-order force constants in the diamond-type structure, considering various cutoff distances $r_c$.
The lattice constant of the diamond structure is denoted by $a$.
}
\label{symfc:Fig-num-basis-cutoff}
\end{figure}

When dealing with a large number of force constant elements or basis sets, it is practical to introduce a cutoff distance to assign zero values to force constant elements beyond this distance.
Figure \ref{symfc:Fig-num-basis-cutoff} illustrates the dependence of the number of basis vectors on the cutoff distance for the diamond-type structure.
The implementation of a cutoff distance significantly reduces the number of basis vectors, thereby alleviating the computational burden associated with basis set construction and force constant estimation.

\subsection{Force constant estimation}

We will present various applications of the current approach implemented in the symfc code for estimating force constants. 
The first set of applications involves calculating lattice thermal conductivities (LTCs) from displacement-force datasets generated using density functional theory (DFT) calculations. 
The second set focuses on calculating LTCs using machine learning potentials (MLPs). 
These applications demonstrate that the force constants can be estimated efficiently with the current approach.
Furthermore, they emphasize the importance of selecting appropriate parameters and provide recommended settings for estimating both second-order and third-order force constants using this approach.

\subsubsection{Computational details}

Supercell force constants were determined by applying standard linear regression to a displacement-force dataset that consisted of $N_s$ supercell structures. 
They were created by introducing random atomic displacements of magnitude $d$ into the original supercell structure with a given supercell size. 
The forces acting on atoms in these structures were then computed using the DFT calculation or polynomial MLPs \cite{PhysRevB.99.214108,doi:10.1063/5.0129045} representing short-range interatomic interactions with a polynomial function of polynomial invariants for arbitrary rotations.
In applications using DFT calculations, force calculations were performed using the plane-wave-basis projector augmented wave method \cite{PAW1} within the Perdew--Burke--Ernzerhof exchange-correlation functional \cite{GGA:PBE96} as implemented in the \textsc{vasp} code \cite{VASP1,VASP2,PAW2}.
The other computational conditions are given in each application.

The LTCs were calculated as the solution of the Peierls--Boltzmann equation under the relaxation time approximation \cite{Peierls-Boltzmann-1929,
Peierls-Quantum-Theory-of-Solids, 
Allen-LTC-2018} using the \textsc{phono3py} code \cite{phono3py,phonopy-phono3py-JPCM}.
To simplify the LTC calculations and measure the precision of force constants, only phonon-phonon scattering was considered to obtain the phonon relaxation times. 
The phonon lifetimes were calculated from supercell force constants as the reciprocals of the imaginary part of the phonon self-energy corresponding to the bubble diagram, which were used as the relaxation times. 
Phonon properties required for the LTC calculations such as phonon group velocities and mode heat capacities were obtained from dynamical matrices derived from second-order supercell force constants. 
Mesh grids for sampling the reciprocal spaces are shown in each application.

\subsubsection{Estimation from DFT datasets}

\begin{figure}[tbp]
\includegraphics[clip,width=\linewidth]{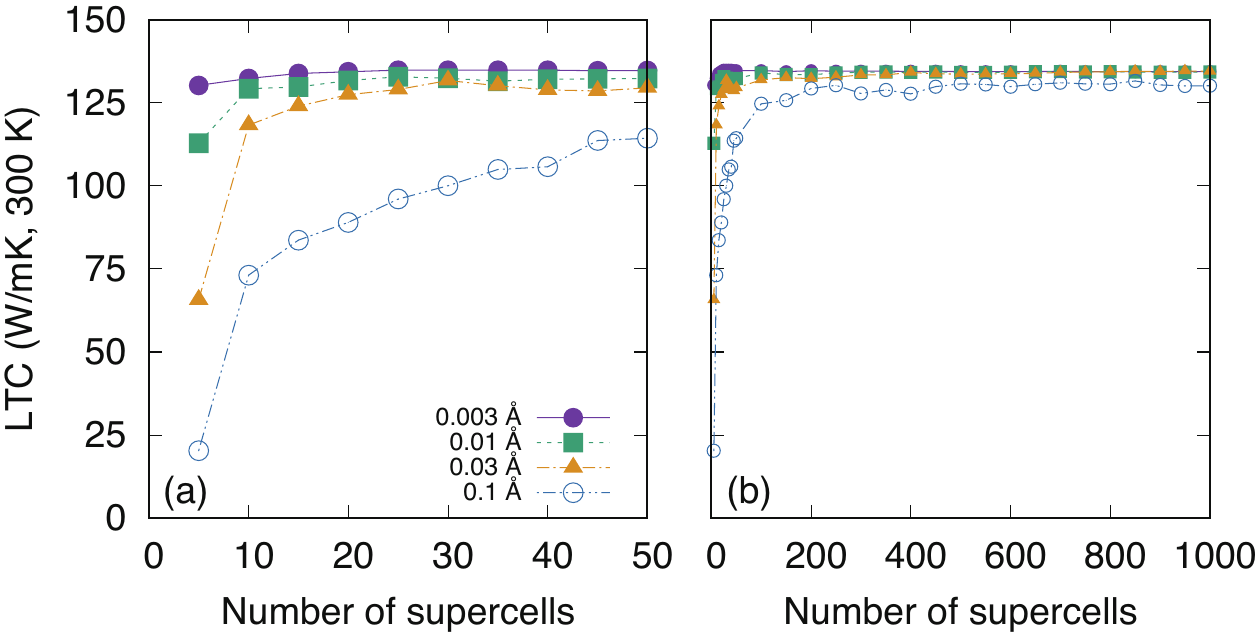}
\caption{
(a) Relationship between the LTC and the number of supercells with displacement magnitudes $d=0.003$, $0.01$, $0.03$, and $0.1$ \AA\ in diamond silicon. 
Displacement-force datasets are derived exclusively from DFT calculations. 
The minimum number of supercells needed to construct a full-rank predictor matrix is five for the $2\times2\times2$ supercell expansion. 
LTC values are shown for up to 50 supercells.
(b) Relationship between LTC and the number of supercells with the same displacement magnitudes in diamond silicon, with LTC values shown for up to 1000 supercells.
}
\label{symfc:Fig-LTC-diamond-Si-DFT}
\end{figure}

Firstly, we demonstrate an application of our approach to diamond silicon, where displacement-force datasets were generated exclusively using DFT calculations. 
In this application, we computed the forces acting on atoms in supercell structures with atomic displacements through DFT calculations and used them to estimate the force constants.
The supercell structures were constructed using the $2\times2\times2$ expansion of the unit cell.
The cutoff energy was set to 300 eV, and the total energies converged to less than 10$^{-3}$ meV/supercell.
We examined atomic displacements with magnitudes of $d=0.003$, $0.01$, $0.03$, and $0.1$ \AA.

In Fig. \ref{symfc:Fig-LTC-diamond-Si-DFT}, we present the relationship between the LTC, the magnitude of atomic displacements, and the number of supercells in the displacement-force dataset. 
When atomic displacements are larger than 0.01 \AA\ and the number of supercells is small, the LTC values are underestimated due to errors in the forces resulting from the truncation of higher-order contributions. 
Consequently, minimal supercell structures are insufficient for accurate LTC prediction.
On the other hand, accurate LTC values can be obtained even using a small dataset with atomic displacements of 0.003 \AA. 
In the diamond structure, the atomic coordinates have no degrees of freedom, which means that no residual forces arise. 
Consequently, numerical and truncation errors in the dataset are minimal.

\begin{figure}[tbp]
\includegraphics[clip,width=\linewidth]{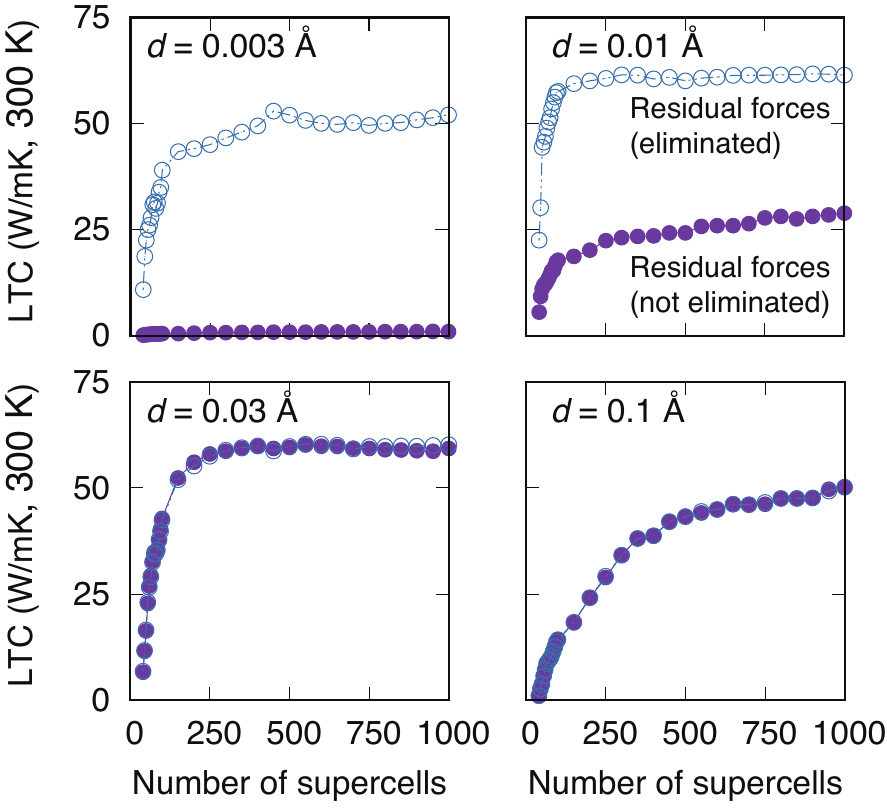}
\caption{
Convergence behavior of LTC with respect to the number of supercells in the wurtzite-type ZnS. 
Displacement-force datasets are derived solely from DFT calculations. 
The supercell size is $3\times3\times2$, with a minimum of 36 supercells required to construct a full-rank predictor matrix. 
Average values of the diagonal elements in LTC tensors are presented.
The closed circles represent LTC values calculated using forces without subtracting residual forces, while the open circles represent LTC values calculated using forces with residual forces removed.
}
\label{symfc:Fig-LTC-wurtzite-ZnS-DFT}
\end{figure}

Then, we demonstrate LTC calculations using the DFT calculation for wurzite-type ZnS.
The wurtzite structure affords a degrees of freedom in atomic coordinates, thus suggesting that the equilibrium structure optimized through DFT calculations may possess residual forces approaching the specified convergence tolerance.
To create a displacement-force dataset, the forces acting on atoms in supercell structures were computed using only DFT calculations.
Supercell structures were constructed using the $3\times3\times2$ expansion of the unit cell.
The cutoff energy was set to 500 eV, and the total energies converged to less than $10^{-3}$ meV/supercell.
The atomic positions in the unit cell were optimized until the residual forces were less than $0.0015$ eV/\AA.
To examine the dependency of LTC on atomic displacements, we investigated magnitudes of $d=0.003$, $0.01$, $0.03$, and $0.1$ \AA.

Figure \ref{symfc:Fig-LTC-wurtzite-ZnS-DFT} illustrates the relationship among LTC, the number of supercells, and the magnitude of atomic displacements in wurtzite-type ZnS. 
LTC values derived from raw displacement-force datasets are compared with those derived from datasets where residual forces are subtracted.
Residual forces can introduce numerical errors in force constant calculations, significantly impacting LTC predictions, especially when atomic displacements and resulting forces are small. 
For atomic displacements of $d = 0.003$ and $0.01$ \AA, LTC values exhibit sensitivity to whether residual forces are subtracted from DFT forces in supercell structures, as depicted in Fig. \ref{symfc:Fig-LTC-wurtzite-ZnS-DFT}.
Conversely, for displacements of $d = 0.03$ and $0.1$ \AA, the forces in supercell structures are sufficiently large that LTC values appear independent of residual force subtraction.

The results depicted in Fig. \ref{symfc:Fig-LTC-wurtzite-ZnS-DFT} highlight the importance of subtracting residual forces for accurately and robustly estimating force constants and LTCs.
However, a noticeable discrepancy is observed in the converged LTC values when using $d = 0.003$ \AA\ compared to $d = 0.01$ and $0.03$ \AA.
This difference likely stems from residual forces and other numerical errors in the DFT calculations being relatively large compared to the magnitude of the forces.
Conversely, for $d = 0.1$ \AA, the significant contributions from higher-order effects lead to slow LTC convergence with respect to the number of supercells due to their truncation.
Based on these findings, it is recommended to use atomic displacements in the range of $d = 0.01$ to $0.03$ \AA\ to obtain reliable LTC values in this case.

In comparison to the diamond structure, the wurtzite structure requires a greater number of supercell structures due to the increased number of basis vectors for the force constants. 
In such cases, adopting an efficient approach for determining force constants proves highly advantageous \cite{10.1063/5.0211296}.
This approach combines the current projector-based method with on-the-fly polynomial MLPs developed from DFT displacement-force datasets.
DFT calculations are necessary only for supercell structures used in MLP development, while force calculations for estimating force constants utilize the MLPs.
Moreover, even with small displacements in force calculations, accurate force constants and LTC values can be achieved owing to the smooth potential energy surface of MLPs.
Consequently, this approach significantly reduces the number of supercell structures needed for DFT calculations, while maintaining accuracy comparable to force constant and LTC calculations obtained from DFT calculations for a much larger set of structures.

\subsubsection{Estimation from MLP datasets}

Here, we perform force constant and LTC calculations using displacement-force datasets obtained from MLPs.
The computational costs associated with MLP force calculations are negligible in force constant and LTC calculations.
We examine the performance of the projector-based procedure for applications that are impractical using only DFT calculations.
MLPs offer several advantages:
(1) They enable the generation of extensive displacement-force datasets with significantly reduced computational demands.
(2) They exihibit smooth potential energy surfaces around equilibrium structures, which allows us to employ small magnitudes of atomic displacements deriving supercell structures.
(3) They facilitate the elimination of residual forces in equilibrium structures through local structure optimization.
% While constructing polynomial MLPs on-the-fly is feasible, our study utilizes existing polynomial MLPs.

% \subsubsection{Diamond Si (MLP)}
Force constant and LTC calculations were firstly conducted using MLP displacement-force datasets in diamond silicon.
The polynomial MLP utilized in this study was adopted from prior research \cite{FUJII2022111137}, which investigated LTCs at silicon grain boundaries through molecular dynamics simulations.
This MLP is also accessible in the Polynomial Machine Learning Potential Repository \cite{MachineLearningPotentialRepository}. 
It was trained on a dataset comprising diverse crystal structures and their derivatives, incorporating random atomic displacements and changes in cell shapes.
For LTC calculations, reciprocal spaces were sampled using a $19 \times 19 \times 19$ mesh grid.

\begin{figure*}[tbp]
\includegraphics[clip,width=0.8\linewidth]{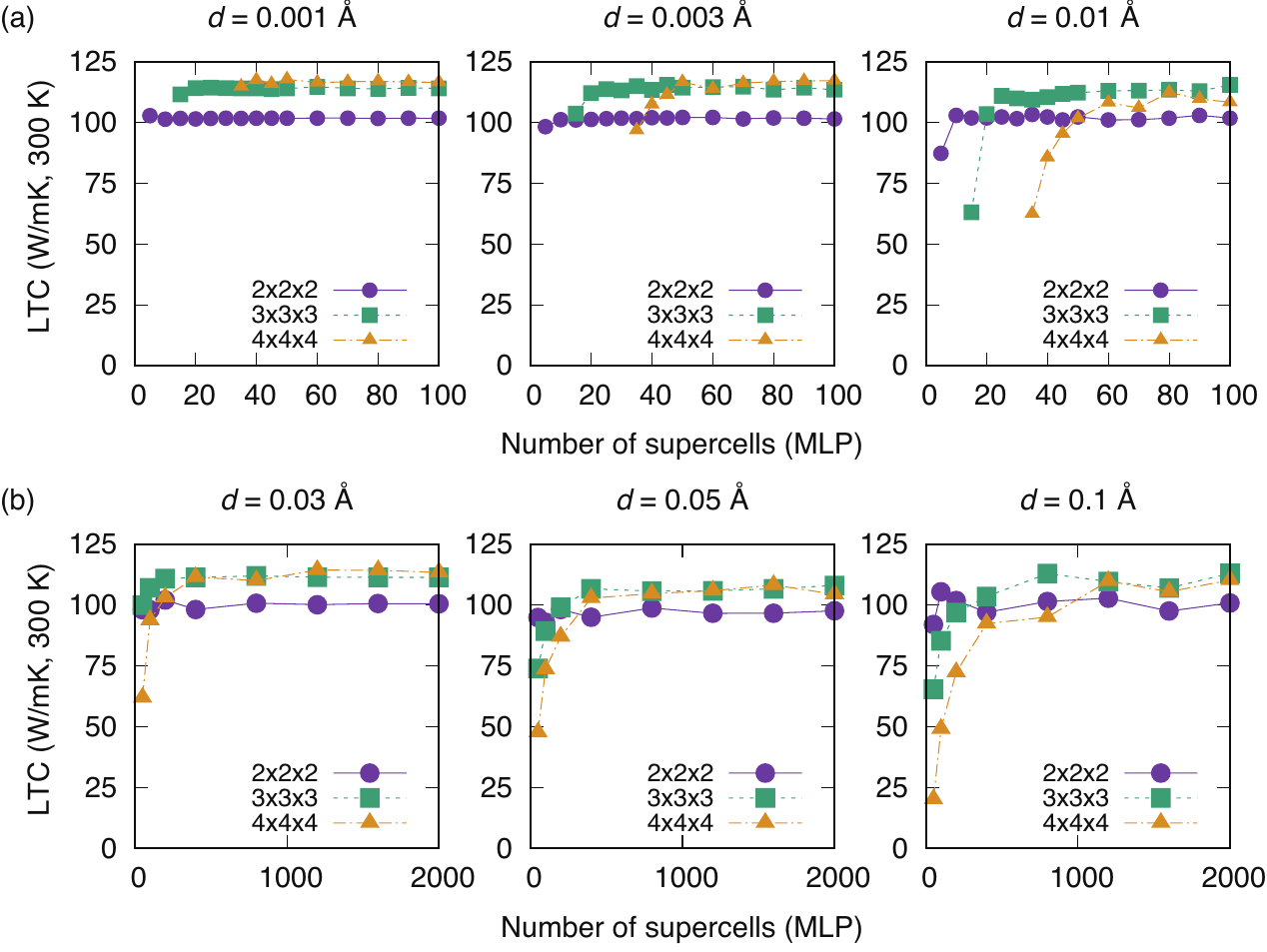}
\caption{
(a) Relationship between the LTC and the number of supercells with displacement magnitudes $d=0.001$, $0.003$, and $0.01$ \AA\ in diamond silicon.
The minimum numbers of supercells required to construct a full-rank predictor matrix are 5, 14, and 33 for the $2\times2\times2$, $3\times3\times3$, and $4\times4\times4$ supercell expansions, respectively.
LTC values are shown for up to 100 supercells.
(b) Relationship between LTC and the number of supercells with displacement magnitudes $d=0.03$, $0.05$, and $0.1$ \AA\ in diamond silicon. 
LTC values are shown for up to 2000 supercells.
}
\label{symfc:Fig-LTC-Si}
\end{figure*}

The current projector-based approach enables efficient estimation of force constants and LTCs for supercells containing up to 512 atoms, when no cutoff distance is applied. 
Figure \ref{symfc:Fig-LTC-Si} illustrates the relationship between the LTC, supercell size, displacement magnitude, and the number of supercell structures within the displacement-force dataset. 
The LTC values obtained from the polynomial MLP using the $2\times2\times2$ supercell are slightly smaller than those obtained from the DFT calculation, as also demonstrated in Ref. \onlinecite{FUJII2022111137}. 
The complete third-order basis set sizes for the $2\times2\times2$, $3\times3\times3$, and $4\times4\times4$ supercell expansions are 777, 8800, and 49301, respectively. 
Consequently, the minimum numbers of supercells required to construct a full-rank predictor matrix $\bm{X}$ are 5, 14, and 33, respectively. 
When considering small displacement values, it is apparent that even a limited number of supercells can yield accurate LTCs. 
Specifically, 5 and 15 supercell structures generated from random displacements with $d = 0.001$ \AA\ are adequate to obtain precise LTCs for the $2\times2\times2$ and $3\times3\times3$ expansions, respectively. 
In contrast, minimal structures fail to provide accurate LTC predictions when larger displacements are used. 
Due to the truncation of significant higher-order contributions, the LTC converges slowly as the number of supercells increases, as evidenced in Fig. \ref{symfc:Fig-LTC-Si} (b).

\begin{figure}[tbp]
\includegraphics[clip,width=0.7\linewidth]{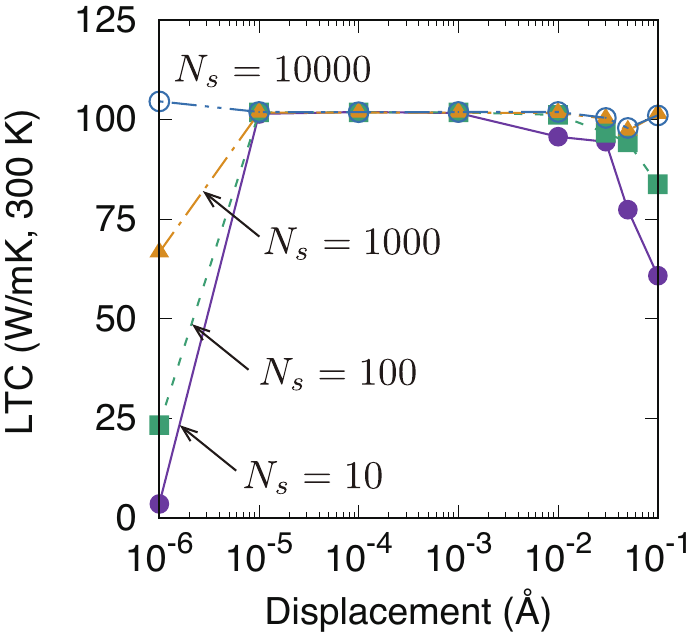}
\caption{
Relationship between the LTC value and the magnitude of atomic displacements in diamond silicon with a supercell size of $2\times2\times2$. 
LTC values are calculated using displacement-force datasets consisting of 10, 100, 1000, and 10000 supercell structures.
}
\label{symfc:Fig-LTC-disp-Si}
\end{figure}

Figure \ref{symfc:Fig-LTC-disp-Si} shows the relationship between the LTC and the magnitude of atomic displacements. 
The LTC values are computed from displacement-force datasets comprising 10, 100, 1000, and 10000 supercell structures. 
For very small displacement values, such as $d=10^{-6}$ \AA, the resulting forces are exceedingly small, which means that numerical noise can significantly affect the accuracy of force constants. 
Consequently, a large number of supercells are required to achieve accurate LTCs. 
To accurately evaluate third-order force constants without the need for renormalizing higher-order contributions, it is preferable to use atomic displacements within the range of $10^{-5}$ to $10^{-3}$ \AA.
Conversely, when atomic displacements exceed 0.01 \AA, a substantial number of supercells are also needed to achieve converged LTC values. 
This requirement arises from truncation errors associated with higher-order contributions, which complicate the computation of converged LTCs.

It is important to note that in this study, we utilized a constant magnitude of atomic displacements to generate each displacement-force dataset. 
However, the current discussion is pertinent to displacement-force datasets that include a range of different magnitudes of atomic displacements. 
Such datasets are typically encountered when estimating temperature-dependent force constants relevant to self-consistent phonons (e.g. \cite{Errea-SSCHA-2013, van-Roekeghem-2020}), which inherently include higher-order contributions to the force constants.
Consequently, when evaluating self-consistent phonons, particularly at high temperatures, a large number of supercell structures will be required.

\begin{figure}[tbp]
\includegraphics[clip,width=0.7\linewidth]{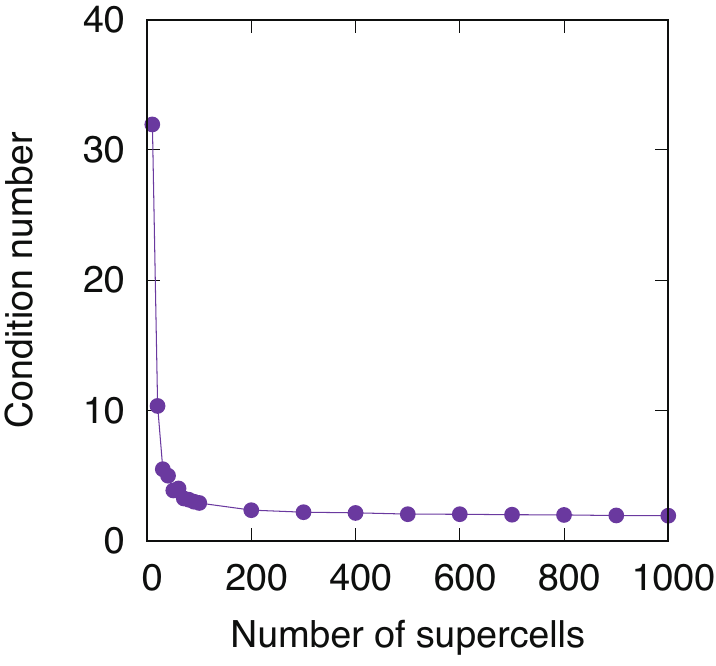}
\caption{
Dependence of the condition number of $\bm{X}^\top \bm{X}$ for third-order force constants on the number of supercells in diamond silicon. 
The supercell size of $2\times2\times2$ and the magnitude of atomic displacements of $d = 0.001$ \AA\ are used.
The condition number is nearly independent of the magnitude of atomic displacements, with similar values observed for $d = 0.1$ \AA. 
}
\label{symfc:Fig-condition-number-Si}
\end{figure}

As described above, the imprecise prediction of LTCs is caused by errors in the forces and is closely related to the linear dependence among predictor variables in matrix $\bm{X}$.
This phenomenon, known as multicollinearity, can render regression coefficients highly sensitive to errors in the forces. 
In standard linear regression, the condition number of matrix $\bm{X}^\top \bm{X}$ is commonly used as an indicator of multicollinearity. 
The condition number is determined by dividing the maximum eigenvalue of $\bm{X}^\top \bm{X}$ by the minimum eigenvalue. 
A significantly large condition number suggests the presence of multicollinearity in the regression analysis.

Figure \ref{symfc:Fig-condition-number-Si} illustrates the dependence of the condition number of matrix $\bm{X}^\top \bm{X}$ for the third-order force constants with the $2\times2\times2$ supercell on the number of supercell structures.
The condition number is high when using only a small number of supercells, close to the minimum number required to construct a full-rank predictor matrix. 
In such cases, the LTC predictions are highly sensitive to force errors, leading to imprecise LTC estimates. 
Conversely, the condition number decreases as the number of supercells increases, and this decrease aligns with the convergence patterns observed in LTCs. 
Therefore, using an orthonormal basis set for force constants along with a sufficiently large number of random displacements for force calculations creates optimal conditions for estimating force constants through standard linear regression.

% \subsubsection{Wurtzite AgI (MLP)}
Next, we calculate LTCs in wurtzite AgI using a polynomial MLP, which was developed from a displacement-force DFT dataset using the on-the-fly approach \cite{10.1063/5.0211296}.
This dataset was made up of 300 supercell structures with atomic displacements of 0.03 \AA.  
We assess the forces acting on atoms for $N_s$ structures using the polynomial MLP. 
Here, the atomic positions optimized by the DFT calculation are used as the equilibrium atomic positions.
Therefore, residual forces at the equilibrium atomic positions are eliminated from the forces to stabilize LTC calculations, particularly with small atomic displacements. 
Additionally, we sample the reciprocal spaces for LTC calculations using the $19 \times 19 \times 10$ mesh.
%Non-analytical term correction~\cite{Pick-1970,Gonze-1994,Gonze-1997} was applied to dynamical matrices to treat long range dipole-dipole interactions for the harmonic phonon calculation. 

\begin{figure}[tbp]
\includegraphics[clip,width=0.7\linewidth]{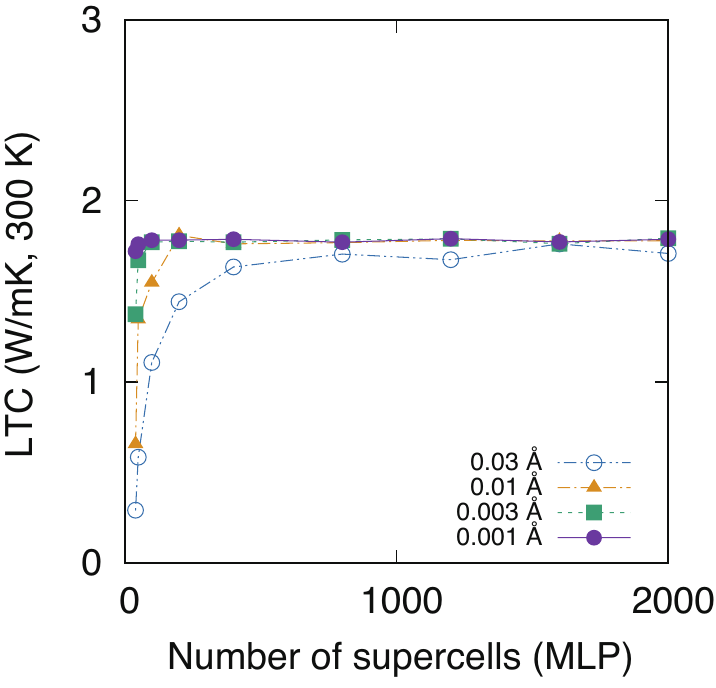}
\caption{
Convergence behavior of LTC with respect to the number of supercells in wurtzite AgI.
The supercell size is $3\times3\times2$, and the number of basis vectors for the third-order force constants is 7752, necessitating a minimum of 36 supercell structures.
The average values of the diagonal elements in LTC tensors are presented.
}
\label{symfc:Fig-LTC-wurtzite-AgI}
\end{figure}

Figure \ref{symfc:Fig-LTC-wurtzite-AgI} shows the relationship between the LTC and the number of supercells in wurtzite AgI. 
The supercells are constructed using the $3\times3\times2$ expansion of the unit cell. 
To construct a full-rank predictor matrix $\bm{X}$, a minimum of 36 supercell structures is required because there are 7752 basis vectors for the third-order force constants. 
The convergence behavior of the LTC is influenced by the magnitude of atomic displacements. 
For example, a small number of supercell structures are sufficient to achieve a converged LTC value for small atomic displacements, analogous to the situation observed with diamond silicon. 
Conversely, for larger displacements, a greater number of supercell structures are necessary to obtain a precise LTC value that effectively incorporates higher-order contributions.

\subsubsection{Estimation from MLP datasets using cutoff distance}

\begin{figure}[tbp]
\includegraphics[clip,width=0.75\linewidth]{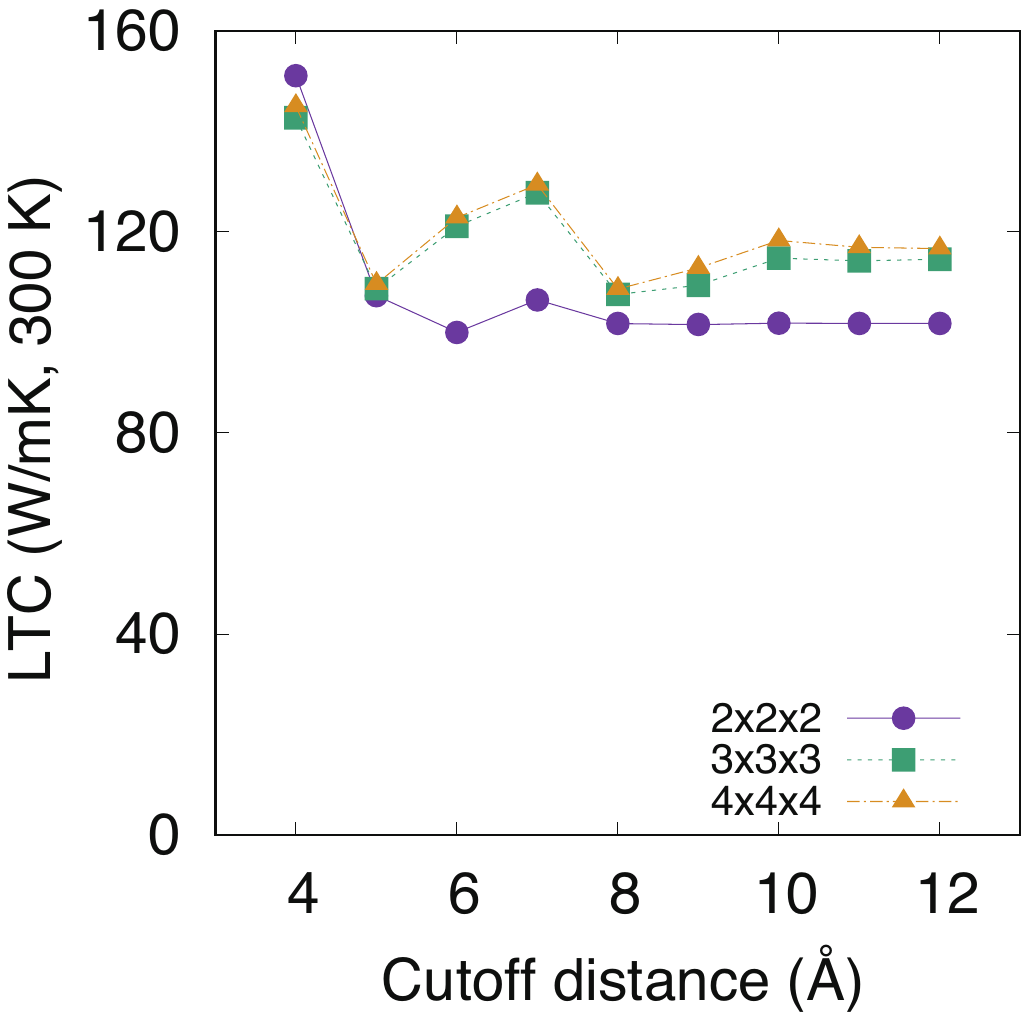}
\caption{
Dependence of the LTC on the cutoff distance in diamond silicon.
The cutoff distances are applied only to the third-order force constants. 
Force calculations for supercell structures are performed using a polynomial MLP with a cutoff radius of 6 \AA. 
The displacement magnitude is set to $d=0.001$ \AA\, and 100 supercell structures are used.
}
\label{symfc:Fig-LTC-Si-cutoff}
\end{figure}

Finally, we investigate the dependence of the LTC on the cutoff distance for third-order force constants in diamond silicon. 
No cutoff distance is applied to the second-order force constants.
Figure \ref{symfc:Fig-LTC-Si-cutoff} shows how the LTC in diamond silicon varies with the cutoff distance.
Force calculations for supercell structures are performed using the polynomial MLP described in the previous section.
The displacement magnitude is set to $d=0.001$ \AA\, and the number of supercell structures is 100. 
Although the polynomial MLP has a cutoff radius of 6 \AA, many-body terms of the polynomial MLP contribute to LTC predictions with cutoff distances exceeding 6 \AA. 
For diamond silicon, a cutoff distance of 8 \AA\ in force constant calculations yields reasonable LTC values.
Thus, the introduction of the cutoff distance significantly reduces the size of the basis set and computational demands while still providing accurate force constant calculations.

\section{Conclusion}
\label{symfc:sec-conclusion}
In this study, we have proposed a projector-based approach for efficiently constructing a complete orthonormal basis set for force constants. 
We have introduced several compression techniques to reduce the basis set size efficiently.
These techniques involve compressing a projection matrix using eigenvectors of another projection matrix onto subspaces defined by partial requirements for the force constants. 
Our efficient procedures for basis set construction and force constant estimation, up to the fourth order, are implemented in the symfc code \cite{symfc-project}.

The obtained basis set meets necessary requirements for the force constants, including permutation symmetry rules, sum rules, and symmetric properties dictated by the space group operations of the target structure.
Consequently, this complete orthonormal basis set facilitates the precise determination of force constants, ensuring that they exactly satisfy the required constraints. 
As demonstrated in the applications of this study, properties related to the force constants, such as LTCs, can be calculated both accurately and efficiently using this basis set.

\begin{acknowledgments}
This work was supported by JSPS KAKENHI Grant Numbers JP22H01756, and JP19H05787, JP24K08021, and JP21K04632.
\end{acknowledgments}

\appendix

\section{Derivation of projection matrix onto vector space}
\label{symfc:sec:derivation-of-projector}
In this section, we derive the orthogonal projection matrix onto \({\rm Ker} (\bm{C}^\top)\) \cite{Strang_2023,Yanai:1414711}.
We begin by deriving the projection matrix onto the vector space spanned by \(\bm{C}\), denoted as \(\bm{P}_C\). 
Assuming that the column vectors of \(\bm{C}\) are linearly independent, the projection of a vector \(\bm{x}\) onto the vector space spanned by \(\bm{C}\) is expressed as
\begin{equation}
\label{symfc:eq:projection-vector}
\bm{P}_C \bm{x} = \bm{C} \bm{\alpha},
\end{equation}
where the projected vector is represented as a linear combination of the column vectors of \(\bm{C}\), with the coefficients denoted by the vector \(\bm{\alpha}\). 
Since the residual vector \(\bm{x} - \bm{C} \bm{\alpha}\) is orthogonal to all column vectors of \(\bm{C}\), we have the relationship:
\begin{equation}
\bm{C}^\top [\bm{x} - \bm{C} \bm{\alpha}] = 0.
\end{equation}
This can be transformed to yield $\bm{C}^\top \bm{C} \bm{\alpha} = \bm{C}^\top \bm{x}$, leading to the expression $\bm{\alpha} = (\bm{C}^\top \bm{C})^{-1} \bm{C}^\top \bm{x}$.
By substituting this expression into Eq. (\ref{symfc:eq:projection-vector}), we derive the projection matrix onto the vector space spanned by \(\bm{C}\) as
\begin{equation}
\bm{P}_C = \bm{C} (\bm{C}^\top \bm{C})^{-1} \bm{C}^\top.
\end{equation}
Since \({\rm Ker} (\bm{C}^\top)\) constitutes the complementary subspace of the vector space spanned by \(\bm{C}\), the orthogonal projection matrix onto \({\rm Ker} (\bm{C}^\top)\) can be expressed as
\begin{equation}
\bm{P}_{{\rm Ker} (\bm{C}^\top)} = \bm{I} - \bm{C} (\bm{C}^\top \bm{C})^{-1} \bm{C}^\top,
\end{equation}
where \(\bm{I}\) denotes the identity matrix.
When $\bm{C}$ is an orthogonal matrix, the orthogonal projection matrix onto \({\rm Ker} (\bm{C}^\top)\) can be represented as
\begin{equation}
\bm{P}_{{\rm Ker} (\bm{C}^\top)} = \bm{I} - \bm{C} \bm{C}^\top.
\end{equation}

\section{Proofs of commutation relations between projection matrices}
\label{symfc:sec:commutativity-proof}
In this section, we provide a proof of the commutation relations between projection matrices, specifically showing that \([\bm{P}_{\rm spg}, \bm{P}_v] = 0\) and \([\bm{P}_{\rm spg}, \bm{P}_w] = 0\). 
Although we only present the proof for second-order force constants, the argument can be extended to third-order force constants as well.
Consider two orthogonal projection matrices, \(\bm{P}_1\) and \(\bm{P}_2\). 
The transpose of the product \(\bm{P}_1 \bm{P}_2\) is given by $(\bm{P}_1 \bm{P}_2)^\top = \bm{P}_2^\top \bm{P}_1^\top = \bm{P}_2 \bm{P}_1$.
Therefore, if \(\bm{P}_1 \bm{P}_2\) is symmetric, i.e., \(\bm{P}_1 \bm{P}_2 = (\bm{P}_1 \bm{P}_2)^\top\), then \(\bm{P}_1\) and \(\bm{P}_2\) commute.
We will now demonstrate that the product of two projection matrices is symmetric.

(1) Proof of $[\bm{P}_{\rm spg}, \bm{P}_v] = 0$.
Firstly, we will show that $\bm{P}_v \bm{P}_{\rm spg}$ is symmetric, $\bm{P}_v \bm{P}_{\rm spg} = (\bm{P}_v \bm{P}_{\rm spg})^\top$.
Since the projection matrix $\bm{P}_v$ is given by $\bm{P}_v = \bm{I} - \bm{C}_v \bm{C}_v^\top$ as shown in Eq. (\ref{symfc:eqn:sum_rules}),
the elements of $\bm{P}_v^c = \bm{C}_v \bm{C}_v^\top$ are expressed as
\begin{equation}
P^c_{v,(i\alpha j\beta, i'\alpha' j'\beta')} = \frac{1}{N} \delta_{ii'} \delta_{\alpha\alpha'} \delta_{\beta\beta'},
\end{equation}
where $\delta_{ii'}$ denotes the Kronecker delta.
Therefore, when denoting the product of $\bm{P}_v^c$ and $\bm{P}_{\rm spg}$ as $\bm{P}'= \bm{P}_v^c \bm{P}_{\rm spg}$, its elements are written as
\begin{eqnarray}
P'_{(i\alpha j\beta, i'\alpha' j'\beta')} 
&=& 
\sum_{k\gamma l\delta} P^c_{v,(i\alpha j\beta, k\gamma l\delta)} 
P_{{\rm spg}{(k\gamma l\delta, i'\alpha' j'\beta')}} \nonumber \\
&=&
\frac{1}{N} 
\sum_{l} P_{{\rm spg}{(i\alpha l\beta, i'\alpha' j'\beta')}}
\end{eqnarray}
and 
\begin{eqnarray}
P'_{(i'\alpha' j'\beta', i\alpha j\beta)} 
&=& 
\sum_{k\gamma l\delta} P^c_{v,(i'\alpha' j'\beta', k\gamma l\delta)} 
P_{{\rm spg}{(k\gamma l\delta, i\alpha j\beta)}} \nonumber \\
&=&
\frac{1}{N} 
\sum_{l} P_{{\rm spg}{(i'\alpha' l\beta', i\alpha j\beta)}} \nonumber \\
&=& \frac{1}{N} 
\sum_l P_{{\rm spg}{(i\alpha j\beta, i'\alpha' l\beta')}}.
\end{eqnarray}
The final equality is derived from the symmetric property of $\bm{P}_{\rm spg}$.
Using the definition of $\bm{P}_{\rm spg}$, these elements are expressed as
\begin{align}
\label{symfc:eq-pv-pspg-1}
& P'_{(i\alpha j\beta, i'\alpha' j'\beta')} \nonumber \\
& = \frac{1}{N|\mathcal{G}|} \sum_l \sum_{\hat g \in \mathcal{G}} 
\Gamma(\hat g)_{(i\alpha,i'\alpha')} \Gamma({\hat g})_{(l\beta,j'\beta')}
\end{align}
and
\begin{align}
\label{symfc:eq-pv-pspg-2}
& P'_{(i'\alpha' j'\beta', i\alpha j\beta)} \nonumber \\
& = \frac{1}{N|\mathcal{G}|} \sum_l \sum_{\hat g \in \mathcal{G}} 
\Gamma(\hat g)_{(i\alpha,i'\alpha')} \Gamma({\hat g})_{(j\beta,l\beta')},
\end{align}
respectively.
The summations over index $l$ in Eqs. (\ref{symfc:eq-pv-pspg-1}) and (\ref{symfc:eq-pv-pspg-2}) must then be equal, i.e., 
\begin{equation}
\sum_l \Gamma({\hat g})_{(l\beta,j'\beta')} = \sum_l \Gamma({\hat g})_{(j\beta,l\beta')},
\end{equation}
because $\Gamma({\hat g})_{(l\beta,j'\beta')}$ can be non-zero only for single atom $l$ such that operation $\hat g$ moves atom $l$ to atom $j'$ and its value corresponds to the representation of $\hat g$ for the Cartesian components $(\beta, \beta')$.
When we denote these summations as $\Gamma'(\hat g)_{(\beta,\beta')}$,
the elements of $\bm{P}'$ are written as
\begin{align}
& P'_{(i\alpha j\beta, i'\alpha' j'\beta')} 
= P'_{(i'\alpha' j'\beta', i\alpha j\beta)} \nonumber \\
& = \frac{1}{N|\mathcal{G}|} \sum_{\hat g \in \mathcal{G}} \Gamma(\hat g)_{(i\alpha,i'\alpha')} \Gamma'(\hat g)_{(\beta,\beta')}, 
\end{align}
which is equivalent to $\bm{P}_v^c \bm{P}_{\rm spg} = (\bm{P}_v^c \bm{P}_{\rm spg})^\top$. 
Therefore, $\bm{P}_v \bm{P}_{\rm spg} = (\bm{P}_v \bm{P}_{\rm spg})^\top$ can also be derived, and it has been proved that $[\bm{P}_{\rm spg}, \bm{P}_v] = 0$.

(2) Proof of $[\bm{P}_{\rm spg}, \bm{P}_w] = 0$.
We demonstrate that $\bm{P}_w \bm{P}_{\rm spg}$ is symmetric.
The projection matrix $\bm{P}_w$ is represented by $\bm{P}_w = \bm{B}_w \bm{B}_w^\top$.
Therefore, the elements of matrix $\bm{P}'' = \bm{P}_w \bm{P}_{\rm spg}$ are written as
\begin{equation}
P''_{(i\alpha j\beta, i'\alpha' j'\beta')} = \frac{1}{2}
\left[ 
P_{{\rm spg}{(i\alpha j\beta, i'\alpha' j'\beta')}}
+ P_{{\rm spg}{(j\beta i\alpha, i'\alpha' j'\beta')}} 
\right]
\end{equation}
and 
\begin{equation}
P''_{(i'\alpha' j'\beta', i\alpha j\beta)} = \frac{1}{2}
\left[ 
P_{{\rm spg}{(i'\alpha' j'\beta', i\alpha j\beta)}}
+ P_{{\rm spg}{(j'\beta' i'\alpha', i\alpha j\beta)}}
\right] .
\end{equation}
Then, the projection matrix $\bm{P}_{\rm spg}$ is orthogonal and has the following symmetric properties of
\begin{equation}
P_{{\rm spg}{(i\alpha j\beta, i'\alpha' j'\beta')}} = 
P_{{\rm spg}{(i'\alpha' j'\beta', i\alpha j\beta)}} 
\end{equation}
and
\begin{eqnarray}
P_{{\rm spg}{(j\beta i\alpha, i'\alpha' j'\beta')}}
&=& P_{{\rm spg}{(i'\alpha' j'\beta', j\beta i\alpha)}} \nonumber \\
&=& P_{{\rm spg}{(j'\beta' i'\alpha', i\alpha j\beta)}} ,
\end{eqnarray}
where the final equality is derived from the fact that $\bm{P}_{\rm spg}$ is expressed as the sum of the Kronecker products of representation $\bm{\Gamma} (\hat R)$.
After applying these equations, we can conclude that $\bm{P}''$ is symmetric as 
\begin{equation}
P''_{(i\alpha j\beta, i'\alpha' j'\beta')} = P''_{(i'\alpha' j'\beta', i\alpha j\beta)},
\end{equation}
which proves that $[\bm{P}_{\rm spg}, \bm{P}_w] = 0$.

\section{Implementation of force-constant estimation}
\label{symfc:sec-appendix-fc-estimation}
In this section, we present a procedure for efficiently computing $\bm{X}^\top \bm{X}$ and $\bm{X}^\top\bm{y}$ when solving the normal equations derived from a displacement-force dataset.
The basis set of the full projection matrix is obtained from two decomposed matrices, as specified in Eq. (\ref{symfc:eqn:full-basis}).
In this context, the matrices $\bm{B}_{{\rm spg} \cap w}$ and $\bm{B}_{v\downarrow({\rm spg} \cap w)}$ appearing in Eq. (\ref{symfc:eqn:full-basis}) are denoted by $\bm{D}$ and $\bm{E}$, respectively. 
Consequently, the basis sets $\bm{B}^{\rm (FC2)}$ and $\bm{B}^{\rm (FC3)}$ are written by
\begin{eqnarray}
\bm{B}^{\rm (FC2)} &=& \bm{D}^{\rm (FC2)} \bm{E}^{\rm (FC2)} \nonumber \\
\bm{B}^{\rm (FC3)} &=& \bm{D}^{\rm (FC3)} \bm{E}^{\rm (FC3)}.
\end{eqnarray}
Using these expressions, the force constants are represented by
\begin{eqnarray}
\Theta_{i\alpha,j\beta} &=& 
\sum_b c_b^{\rm (FC2)} B_{i\alpha,j\beta,b}^{\rm (FC2)} \nonumber \\
&=& \sum_b c_b^{\rm (FC2)} \left[ \sum_d D_{i\alpha,j\beta,d}^{\rm (FC2)} E_{d,b}^{\rm (FC2)} \right]
\end{eqnarray}
and
\begin{eqnarray}
\Theta_{i\alpha,j\beta,k\gamma} &=& 
\sum_b c_b^{\rm (FC3)} B_{i\alpha,j\beta,k\gamma,b}^{\rm (FC3)} \nonumber \\
&=& \sum_b c_b^{\rm (FC3)} \left[ \sum_d D_{i\alpha,j\beta,k\gamma,d}^{\rm (FC3)} E_{d,b}^{\rm (FC3)} \right],
\end{eqnarray}
where $c_b$ denotes the expansion coefficient for $b$-th basis.
We introduce the variables $z_{i\alpha,d}^{(s,{\rm FC2})}$ and $z_{i\alpha,d}^{(s,{\rm FC3})}$ defined by
\begin{eqnarray}
z_{i\alpha,d}^{(s,{\rm FC2})} &=& 
- \sum_{j\beta} u_{j\beta}^{(s)} D_{i\alpha,j\beta,d}^{\rm (FC2)} 
\nonumber \\
z_{i\alpha,d}^{(s,{\rm FC3})} &=& 
- \frac{1}{2} \sum_{j\beta} \sum_{k\gamma} u_{j\beta}^{(s)} u_{k\gamma}^{(s)} D_{i\alpha,j\beta,k\gamma,d}^{\rm (FC3)}.
\end{eqnarray}
The forces acting on atoms in the supercell structure $s$ can then be expressed as
\begin{eqnarray}
f_{i\alpha}^{(s)} &=&
\sum_b c_b^{\rm (FC2)} \sum_d z_{i\alpha,d}^{(s,{\rm FC2})} E_{d,b}^{\rm (FC2)} \nonumber \\
&+& \sum_b c_b^{\rm (FC3)} \sum_d z_{i\alpha,d}^{(s,{\rm FC3})} E_{d,b}^{\rm (FC3)}.
\end{eqnarray}
In matrix form, the forces are represented as
\begin{eqnarray}
\bm{f}^{(s)} &=& \bm{Z}^{(s)} \bm{E} \bm{c} \nonumber\\
             &=& \bm{X}^{(s)} \bm{c},
\end{eqnarray}
where
\begin{eqnarray}
  \bm{Z}^{(s)} &=&
  \begin{bmatrix}
  \bm{Z}^{(s, \rm FC2)} & \bm{Z}^{(s, \rm FC3)} \\
  \end{bmatrix} 
  \nonumber \\
  &=&
  \begin{bmatrix}
  \cdots & \vdots & \cdots & \vdots & \cdots \\
  \cdots & z_{i\alpha,d}^{(s,{\rm FC2})} &\cdots& z_{i\alpha,d}^{(s,{\rm FC3})} & \cdots\\
  \cdots & \vdots & \cdots & \vdots & \cdots \\
  \end{bmatrix}, 
\end{eqnarray}
and
\begin{equation}
  \bm{E} = 
  \begin{bmatrix}
  \bm{E}^{(\rm FC2)} & \bm{0}\\
  \bm{0} & \bm{E}^{(\rm FC3)} \\
  \end{bmatrix}.
\end{equation}
For the displacement-force dataset, the matrix $\bm{Z}$ is expressed as
\begin{eqnarray}
  \bm{Z} &=& 
  \begin{bmatrix}
  \bm{Z}^{(1)} \\
  \bm{Z}^{(2)} \\
  \bm{Z}^{(3)} \\
  \vdots \\
  \bm{Z}^{(N_s)}
  \end{bmatrix}
  = 
  \begin{bmatrix}
  \bm{Z}^{(1,\rm FC2)} & \bm{Z}^{(1,\rm FC3)} \\
  \bm{Z}^{(2,\rm FC2)} & \bm{Z}^{(2,\rm FC3)} \\
  \bm{Z}^{(3,\rm FC2)} & \bm{Z}^{(3,\rm FC3)} \\
  \vdots & \vdots\\
  \bm{Z}^{(N_s,\rm FC2)} & \bm{Z}^{(N_s,\rm FC3)} \\
  \end{bmatrix} \nonumber \\
  &=& 
  \begin{bmatrix}
  \bm{Z}^{(\rm FC2)} & \bm{Z}^{(\rm FC3)} \\
  \end{bmatrix}.
\end{eqnarray}
Using these representations, $\bm{X}^\top \bm{X}$ and $\bm{X}^\top\bm{y}$ can be rewritten as
\begin{widetext}
\begin{eqnarray}
\bm{X}^\top \bm{X} &=& \bm{E}^\top \bm{Z}^\top \bm{Z} \bm{E}
  =
  \begin{bmatrix}
  \bm{E}^{(\rm FC2)\top} \bm{Z}^{(\rm FC2)\top} \bm{Z}^{(\rm FC2)} \bm{E}^{(\rm FC2)} & \bm{E}^{(\rm FC2)\top} \bm{Z}^{(\rm FC2)\top} \bm{Z}^{(\rm FC3)} \bm{E}^{(\rm FC3)} \\
  \bm{E}^{(\rm FC3)\top} \bm{Z}^{(\rm FC3)\top} \bm{Z}^{(\rm FC2)} \bm{E}^{(\rm FC2)} & \bm{E}^{(\rm FC3)\top} \bm{Z}^{(\rm FC3)\top} \bm{Z}^{(\rm FC3)} \bm{E}^{(\rm FC3)} \\
  \end{bmatrix},
\end{eqnarray}
\end{widetext}
and
\begin{eqnarray}
\bm{X}^\top \bm{y} &=& \bm{E}^\top \bm{Z}^\top \bm{y}
  =
  \begin{bmatrix}
  \bm{E}^{(\rm FC2)\top} \bm{Z}^{(\rm FC2)\top} \bm{y} \\
  \bm{E}^{(\rm FC3)\top} \bm{Z}^{(\rm FC3)\top} \bm{y} \\
  \end{bmatrix},
\end{eqnarray}
respectively.
The matrix products with forms of $\bm{Z}^\top \bm{Z}$ and $\bm{Z}^\top \bm{y}$ are computed efficiently by partitioning $\bm{Z}$ and $\bm{y}$ into subsets, and $\bm{X}^\top \bm{X}$ and $\bm{X}^\top \bm{y}$ are calculated using $\bm{Z}^\top \bm{Z}$ and $\bm{Z}^\top \bm{y}$.

\section{Computational efficiency}
\label{symfc:sec:comptational-efficiency}

\begin{figure*}[tbp]
\includegraphics[clip,width=0.7\linewidth]{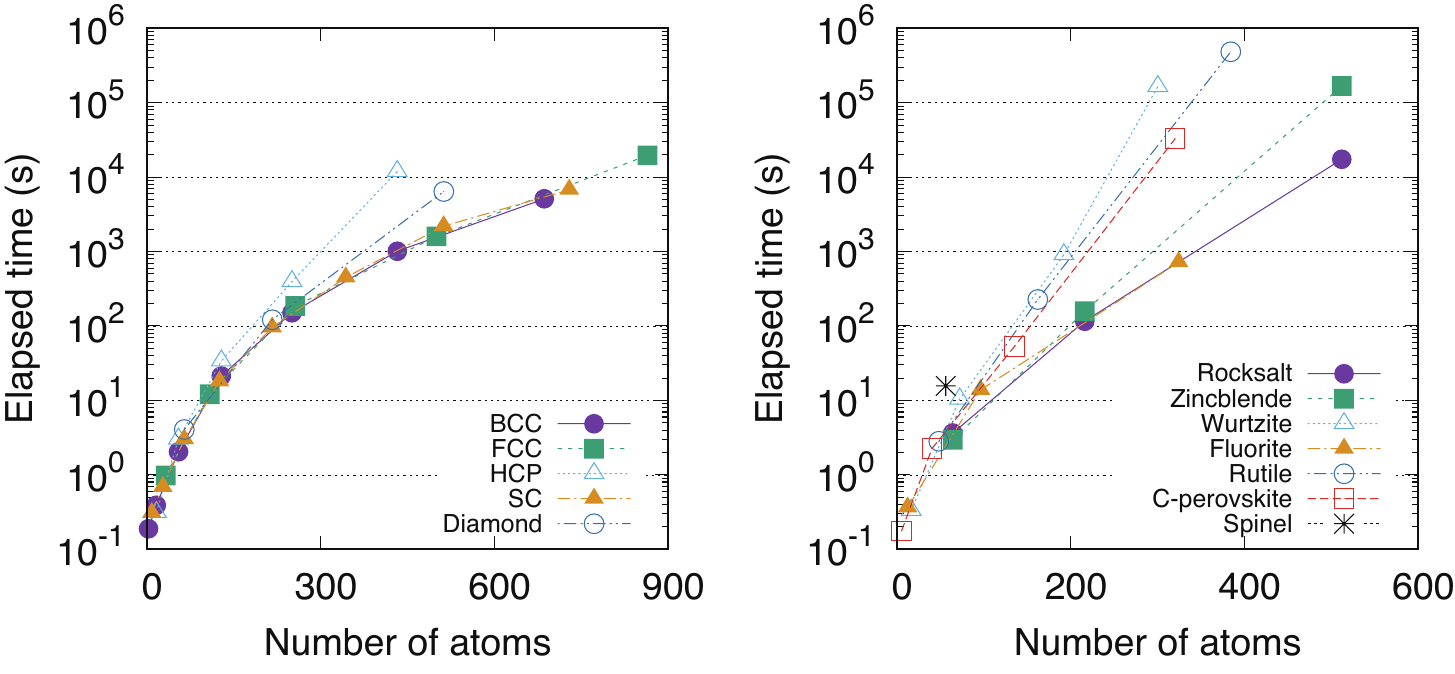}
\caption{
Elapsed time required for computing basis sets of second-order and third-order supercell force constants for elemental and ionic structures.
Elapsed times were measured using the Intel(R) Xeon(R) Gold 6240 CPU.
}
\label{symfc:Fig-time-basis}
\end{figure*}

In this section, we present benchmark results that illustrate the computational efficiency of generating force-constant basis sets and computing force constants from displacement-force datasets. 
Figure \ref{symfc:Fig-time-basis} depicts the elapsed time required for computing the basis sets of second-order and third-order supercell force constants for both elemental and ionic structures.
The computational cost primarily depends on the number of atoms in the supercell, as the number of elements in the second-order and third-order force-constant matrices scales with $9N^2$ and $27N^3$, respectively. 
Additionally, the computational cost varies according to the symmetry of the target structure. 
Structures with lower symmetry generally need higher computational costs compared to those with higher symmetry. 
This is because lower-symmetry structures lack the symmetric properties that could efficiently reduce the size of the basis sets. 
Consequently, solving the compressed eigenvalue problem as described by Eq. (\ref{symfc:eq-procedure-eig-sum}) becomes more computationally demanding for these structures.

\begin{figure*}[tbp]
\includegraphics[clip,width=0.8\linewidth]{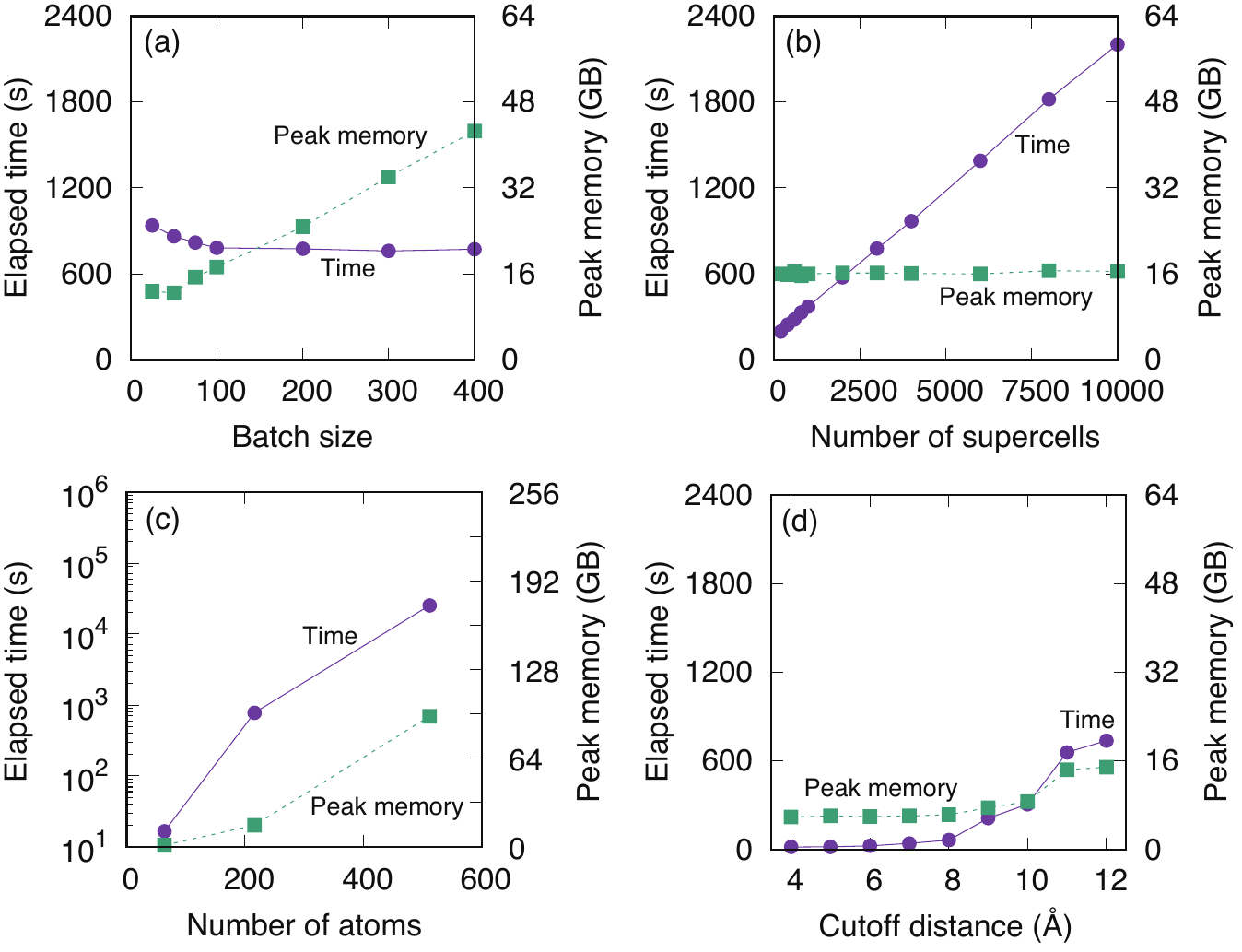}
\caption{
(a) Elapsed time and peak memory usage for estimating second- and third-order force constants with varying subset sizes in diamond silicon. The number of supercells is 3000, with a supercell size of \(3 \times 3 \times 3\) (216 atoms), and atomic displacements are set to \(d = 0.001\) \AA.
(b) Elapsed time and peak memory usage for estimating second- and third-order force constants with a fixed subset size of 100 and varying numbers of supercells in diamond silicon. 
The supercell size is \(3 \times 3 \times 3\) (216 atoms), and the magnitude of atomic displacements is \(d = 0.001\) \AA.
(c) Elapsed time and peak memory usage for estimating second- and third-order force constants with a fixed subset size of 100 and a fixed number of supercells (3000) in diamond silicon. 
The magnitude of atomic displacements is \(d = 0.001\) \AA.
(d) Dependence of elapsed time and peak memory usage on the cutoff distance for estimating second- and third-order force constants with a fixed subset size of 100, a fixed number of supercells (3000), and atomic displacements of \(d = 0.001\) \AA. 
Elapsed times were measured using the Intel(R) Xeon(R) Gold 6240 CPU.
}
\label{symfc:Fig-efficiency-Si}
\end{figure*}

We conducted a benchmark to estimate second- and third-order force constants from displacement-force datasets using a polynomial MLP in diamond silicon. 
Figure \ref{symfc:Fig-efficiency-Si} (a) illustrates the elapsed time and peak memory usage associated with estimating force constants for various subset sizes. 
For this benchmark, we utilized a dataset consisting of 3000 structures with a supercell size of $3\times3\times3$ in diamond silicon. 
As depicted in Fig. \ref{symfc:Fig-efficiency-Si} (a), the computational cost remains relatively constant for subset sizes up to 400 structures, whereas the memory requirements increase approximately proportionally with the subset size. 
Consequently, we recommend that the subset size should not exceed 100 to optimize performance.

Figure \ref{symfc:Fig-efficiency-Si} (b) displays the elapsed time and peak memory usage for estimating force constants with a fixed subset size of 100 but varying numbers of supercells. 
The computational cost scales linearly with the number of supercells in the dataset, while the memory usage remains nearly constant, regardless of the number of supercells. 
This result demonstrates that partitioning $\bm{X}^\top \bm{X}$ effectively mitigates memory consumption.

In Fig. \ref{symfc:Fig-efficiency-Si} (c), we present the elapsed time and peak memory usage for estimating force constants with a fixed subset size of 100 and a fixed number of supercells (3000).
We used three different supercell sizes: $2\times2\times2$, $3\times3\times3$, and $4\times4\times4$.
The results indicate that both computational cost and memory requirements are significantly influenced by the number of atoms in the supercell.

Figure \ref{symfc:Fig-efficiency-Si} (d) illustrates the elapsed time and peak memory usage as a function of the cutoff distance. 
For this analysis, we used a subset size of 100 and 3000 supercells. 
Introducing a cutoff distance for the force constants helps reduce computational demands.

\bibliography{symfc}

\end{document}